\pdfoutput=1
\documentclass[11pt,twoside,a4paper,cmspaper,final,collab]{cms-tdr}
\def\svnVersion{28871:28876}\def\cmsCernNo{2010-055}\def\cmsCernDate{\today}\def\cmsMessage{Submitted to Physical Review D}

\begin{document}\cmsNoteHeader{BPH-10-003}
\hyphenation{env-iron-men-tal}
\hyphenation{had-ron-i-za-tion}
\hyphenation{cal-or-i-me-ter}
\hyphenation{de-vices}
\RCS$HeadURL: svn+ssh://svn.cern.ch/reps/tdr2/notes/BPH-10-003/trunk/BPH-10-003.tex $
\RCS$Id: BPH-10-003.tex 15715 2010-07-22 05:52:02Z ishipsey $

%
%
%

\providecommand {\etal}{\mbox{et al.}\xspace} 
\providecommand {\ie}{\mbox{i.e.}\xspace}     
\providecommand {\eg}{\mbox{e.g.}\xspace}     
\providecommand {\etc}{\mbox{etc.}\xspace}     
\providecommand {\vs}{\mbox{\sl vs.}\xspace}      
\providecommand {\mdash}{\ensuremath{\mathrm{-}}} 

\providecommand {\Lone}{Level-1\xspace} 
\providecommand {\Ltwo}{Level-2\xspace}
\providecommand {\Lthree}{Level-3\xspace}

\providecommand{\ACERMC} {\textsc{AcerMC}\xspace}
\providecommand{\ALPGEN} {{\textsc{alpgen}}\xspace}
\providecommand{\CHARYBDIS} {{\textsc{charybdis}}\xspace}
\providecommand{\CMKIN} {\textsc{cmkin}\xspace}
\providecommand{\CMSIM} {{\textsc{cmsim}}\xspace}
\providecommand{\CMSSW} {{\textsc{cmssw}}\xspace}
\providecommand{\COBRA} {{\textsc{cobra}}\xspace}
\providecommand{\COCOA} {{\textsc{cocoa}}\xspace}
\providecommand{\COMPHEP} {\textsc{CompHEP}\xspace}
\providecommand{\EVTGEN} {{\textsc{evtgen}}\xspace}
\providecommand{\FAMOS} {{\textsc{famos}}\xspace}
\providecommand{\GARCON} {\textsc{garcon}\xspace}
\providecommand{\GARFIELD} {{\textsc{garfield}}\xspace}
\providecommand{\GEANE} {{\textsc{geane}}\xspace}
\providecommand{\GEANTfour} {{\textsc{geant4}}\xspace}
\providecommand{\GEANTthree} {{\textsc{geant3}}\xspace}
\providecommand{\GEANT} {{\textsc{geant}}\xspace}
\providecommand{\HDECAY} {\textsc{hdecay}\xspace}
\providecommand{\HERWIG} {{\textsc{herwig}}\xspace}
\providecommand{\HIGLU} {{\textsc{higlu}}\xspace}
\providecommand{\HIJING} {{\textsc{hijing}}\xspace}
\providecommand{\IGUANA} {\textsc{iguana}\xspace}
\providecommand{\ISAJET} {{\textsc{isajet}}\xspace}
\providecommand{\ISAPYTHIA} {{\textsc{isapythia}}\xspace}
\providecommand{\ISASUGRA} {{\textsc{isasugra}}\xspace}
\providecommand{\ISASUSY} {{\textsc{isasusy}}\xspace}
\providecommand{\ISAWIG} {{\textsc{isawig}}\xspace}
\providecommand{\MADGRAPH} {\textsc{MadGraph}\xspace}
\providecommand{\MCATNLO} {\textsc{mc@nlo}\xspace}
\providecommand{\MCFM} {\textsc{mcfm}\xspace}
\providecommand{\MILLEPEDE} {{\textsc{millepede}}\xspace}
\providecommand{\ORCA} {{\textsc{orca}}\xspace}
\providecommand{\OSCAR} {{\textsc{oscar}}\xspace}
\providecommand{\PHOTOS} {\textsc{photos}\xspace}
\providecommand{\PROSPINO} {\textsc{prospino}\xspace}
\providecommand{\PYTHIA} {{\textsc{pythia}}\xspace}
\providecommand{\SHERPA} {{\textsc{sherpa}}\xspace}
\providecommand{\TAUOLA} {\textsc{tauola}\xspace}
\providecommand{\TOPREX} {\textsc{TopReX}\xspace}
\providecommand{\XDAQ} {{\textsc{xdaq}}\xspace}

\providecommand {\DZERO}{D\O\xspace}     


\providecommand{\de}{\ensuremath{^\circ}}
\providecommand{\ten}[1]{\ensuremath{\times \text{10}^\text{#1}}}
\providecommand{\unit}[1]{\ensuremath{\text{\,#1}}\xspace}
\providecommand{\mum}{\ensuremath{\,\mu\text{m}}\xspace}
\providecommand{\micron}{\ensuremath{\,\mu\text{m}}\xspace}
\providecommand{\cm}{\ensuremath{\,\text{cm}}\xspace}
\providecommand{\mm}{\ensuremath{\,\text{mm}}\xspace}
\providecommand{\mus}{\ensuremath{\,\mu\text{s}}\xspace}
\providecommand{\keV}{\ensuremath{\,\text{ke\hspace{-.08em}V}}\xspace}
\providecommand{\MeV}{\ensuremath{\,\text{Me\hspace{-.08em}V}}\xspace}
\providecommand{\GeV}{\ensuremath{\,\text{Ge\hspace{-.08em}V}}\xspace}
\providecommand{\gev}{\GeV}
\providecommand{\TeV}{\ensuremath{\,\text{Te\hspace{-.08em}V}}\xspace}
\providecommand{\PeV}{\ensuremath{\,\text{Pe\hspace{-.08em}V}}\xspace}
\providecommand{\keVc}{\ensuremath{{\,\text{ke\hspace{-.08em}V\hspace{-0.16em}/\hspace{-0.08em}}c}}\xspace}
\providecommand{\MeVc}{\ensuremath{{\,\text{Me\hspace{-.08em}V\hspace{-0.16em}/\hspace{-0.08em}}c}}\xspace}
\providecommand{\GeVc}{\ensuremath{{\,\text{Ge\hspace{-.08em}V\hspace{-0.16em}/\hspace{-0.08em}}c}}\xspace}
\providecommand{\TeVc}{\ensuremath{{\,\text{Te\hspace{-.08em}V\hspace{-0.16em}/\hspace{-0.08em}}c}}\xspace}
\providecommand{\keVcc}{\ensuremath{{\,\text{ke\hspace{-.08em}V\hspace{-0.16em}/\hspace{-0.08em}}c^\text{2}}}\xspace}
\providecommand{\MeVcc}{\ensuremath{{\,\text{Me\hspace{-.08em}V\hspace{-0.16em}/\hspace{-0.08em}}c^\text{2}}}\xspace}
\providecommand{\GeVcc}{\ensuremath{{\,\text{Ge\hspace{-.08em}V\hspace{-0.16em}/\hspace{-0.08em}}c^\text{2}}}\xspace}
\providecommand{\TeVcc}{\ensuremath{{\,\text{Te\hspace{-.08em}V\hspace{-0.16em}/\hspace{-0.08em}}c^\text{2}}}\xspace}

\providecommand{\pbinv} {\mbox{\ensuremath{\,\text{pb}^\text{$-$1}}}\xspace}
\providecommand{\fbinv} {\mbox{\ensuremath{\,\text{fb}^\text{$-$1}}}\xspace}
\providecommand{\nbinv} {\mbox{\ensuremath{\,\text{nb}^\text{$-$1}}}\xspace}
\providecommand{\percms}{\ensuremath{\,\text{cm}^\text{$-$2}\,\text{s}^\text{$-$1}}\xspace}
\providecommand{\lumi}{\ensuremath{\mathcal{L}}\xspace}
\providecommand{\Lumi}{\ensuremath{\mathcal{L}}\xspace}
%
\providecommand{\LvLow}  {\ensuremath{\mathcal{L}=\text{10}^\text{32}\,\text{cm}^\text{$-$2}\,\text{s}^\text{$-$1}}\xspace}
\providecommand{\LLow}   {\ensuremath{\mathcal{L}=\text{10}^\text{33}\,\text{cm}^\text{$-$2}\,\text{s}^\text{$-$1}}\xspace}
\providecommand{\lowlumi}{\ensuremath{\mathcal{L}=\text{2}\times \text{10}^\text{33}\,\text{cm}^\text{$-$2}\,\text{s}^\text{$-$1}}\xspace}
\providecommand{\LMed}   {\ensuremath{\mathcal{L}=\text{2}\times \text{10}^\text{33}\,\text{cm}^\text{$-$2}\,\text{s}^\text{$-$1}}\xspace}
\providecommand{\LHigh}  {\ensuremath{\mathcal{L}=\text{10}^\text{34}\,\text{cm}^\text{$-$2}\,\text{s}^\text{$-$1}}\xspace}
\providecommand{\hilumi} {\ensuremath{\mathcal{L}=\text{10}^\text{34}\,\text{cm}^\text{$-$2}\,\text{s}^\text{$-$1}}\xspace}


\providecommand{\PT}{\ensuremath{p_{\mathrm{T}}}\xspace}
\providecommand{\pt}{\ensuremath{p_{\mathrm{T}}}\xspace}
\providecommand{\ET}{\ensuremath{E_{\mathrm{T}}}\xspace}
\providecommand{\HT}{\ensuremath{H_{\mathrm{T}}}\xspace}
\providecommand{\et}{\ensuremath{E_{\mathrm{T}}}\xspace}
\providecommand{\Em}{\ensuremath{E\hspace{-0.6em}/}\xspace}
\providecommand{\Pm}{\ensuremath{p\hspace{-0.5em}/}\xspace}
\providecommand{\PTm}{\ensuremath{{p}_\mathrm{T}\hspace{-1.02em}/}\xspace}
\providecommand{\PTslash}{\ensuremath{{p}_\mathrm{T}\hspace{-1.02em}/}\xspace}
\providecommand{\ETm}{\ensuremath{E_{\mathrm{T}}^{\text{miss}}}\xspace}
\providecommand{\ETslash}{\ensuremath{E_{\mathrm{T}}\hspace{-1.1em}/}\xspace}
\providecommand{\MET}{\ensuremath{E_{\mathrm{T}}^{\text{miss}}}\xspace}
\providecommand{\ETmiss}{\ensuremath{E_{\mathrm{T}}^{\text{miss}}}\xspace}
\providecommand{\VEtmiss}{\ensuremath{{\vec E}_{\mathrm{T}}^{\text{miss}}}\xspace}

\providecommand{\dd}[2]{\ensuremath{\frac{\mathrm{d} #1}{\mathrm{d} #2}}}

\ifthenelse{\boolean{cms@italic}}{\newcommand{\cmsSymbolFace}{\relax}}{\newcommand{\cmsSymbolFace}{\mathrm}}

\providecommand{\zp}{\ensuremath{\cmsSymbolFace{Z}^\prime}\xspace}
\providecommand{\JPsi}{\ensuremath{\cmsSymbolFace{J}\hspace{-.08em}/\hspace{-.14em}\psi}\xspace}
\providecommand{\Z}{\ensuremath{\cmsSymbolFace{Z}}\xspace}
\providecommand{\ttbar}{\ensuremath{\cmsSymbolFace{t}\overline{\cmsSymbolFace{t}}}\xspace}

\newcommand{\cPgn}{\ensuremath{\nu}}
\newcommand{\cPJgy}{\JPsi}
\newcommand{\cPZ}{\Z}
\newcommand{\cPZpr}{\zp}


\providecommand{\AFB}{\ensuremath{A_\text{FB}}\xspace}
\providecommand{\wangle}{\ensuremath{\sin^{2}\theta_{\text{eff}}^\text{lept}(M^2_\Z)}\xspace}
\providecommand{\stat}{\ensuremath{\,\text{(stat.)}}\xspace}
\providecommand{\syst}{\ensuremath{\,\text{(syst.)}}\xspace}
\providecommand{\kt}{\ensuremath{k_{\mathrm{T}}}\xspace}

\providecommand{\BC}{\ensuremath{\mathrm{B_{c}}}\xspace}
\providecommand{\bbarc}{\ensuremath{\mathrm{\overline{b}c}}\xspace}
\providecommand{\bbbar}{\ensuremath{\mathrm{b\overline{b}}}\xspace}
\providecommand{\ccbar}{\ensuremath{\mathrm{c\overline{c}}}\xspace}
\providecommand{\bspsiphi}{\ensuremath{\mathrm{B_s} \to \JPsi\, \phi}\xspace}
\providecommand{\EE}{\ensuremath{\mathrm{e^+e^-}}\xspace}
\providecommand{\MM}{\ensuremath{\mu^+\mu^-}\xspace}
\providecommand{\TT}{\ensuremath{\tau^+\tau^-}\xspace}

\providecommand{\HGG}{\ensuremath{\mathrm{H}\to\gamma\gamma}}
\providecommand{\GAMJET}{\ensuremath{\gamma + \text{jet}}}
\providecommand{\PPTOJETS}{\ensuremath{\mathrm{pp}\to\text{jets}}}
\providecommand{\PPTOGG}{\ensuremath{\mathrm{pp}\to\gamma\gamma}}
\providecommand{\PPTOGAMJET}{\ensuremath{\mathrm{pp}\to\gamma + \mathrm{jet}}}
\providecommand{\MH}{\ensuremath{M_{\mathrm{H}}}}
\providecommand{\RNINE}{\ensuremath{R_\mathrm{9}}}
\providecommand{\DR}{\ensuremath{\Delta R}}

%

\providecommand{\ga}{\ensuremath{\gtrsim}}
\providecommand{\la}{\ensuremath{\lesssim}}
\providecommand{\swsq}{\ensuremath{\sin^2\theta_\mathrm{W}}\xspace}
\providecommand{\cwsq}{\ensuremath{\cos^2\theta_\mathrm{W}}\xspace}
\providecommand{\tanb}{\ensuremath{\tan\beta}\xspace}
\providecommand{\tanbsq}{\ensuremath{\tan^{2}\beta}\xspace}
\providecommand{\sidb}{\ensuremath{\sin 2\beta}\xspace}
\providecommand{\alpS}{\ensuremath{\alpha_S}\xspace}
\providecommand{\alpt}{\ensuremath{\tilde{\alpha}}\xspace}

\providecommand{\QL}{\ensuremath{\mathrm{Q}_\mathrm{L}}\xspace}
\providecommand{\sQ}{\ensuremath{\tilde{\mathrm{Q}}}\xspace}
\providecommand{\sQL}{\ensuremath{\tilde{\mathrm{Q}}_\mathrm{L}}\xspace}
\providecommand{\ULC}{\ensuremath{\mathrm{U}_\mathrm{L}^\mathrm{C}}\xspace}
\providecommand{\sUC}{\ensuremath{\tilde{\mathrm{U}}^\mathrm{C}}\xspace}
\providecommand{\sULC}{\ensuremath{\tilde{\mathrm{U}}_\mathrm{L}^\mathrm{C}}\xspace}
\providecommand{\DLC}{\ensuremath{\mathrm{D}_\mathrm{L}^\mathrm{C}}\xspace}
\providecommand{\sDC}{\ensuremath{\tilde{\mathrm{D}}^\mathrm{C}}\xspace}
\providecommand{\sDLC}{\ensuremath{\tilde{\mathrm{D}}_\mathrm{L}^\mathrm{C}}\xspace}
\providecommand{\LL}{\ensuremath{\mathrm{L}_\mathrm{L}}\xspace}
\providecommand{\sL}{\ensuremath{\tilde{\mathrm{L}}}\xspace}
\providecommand{\sLL}{\ensuremath{\tilde{\mathrm{L}}_\mathrm{L}}\xspace}
\providecommand{\ELC}{\ensuremath{\mathrm{E}_\mathrm{L}^\mathrm{C}}\xspace}
\providecommand{\sEC}{\ensuremath{\tilde{\mathrm{E}}^\mathrm{C}}\xspace}
\providecommand{\sELC}{\ensuremath{\tilde{\mathrm{E}}_\mathrm{L}^\mathrm{C}}\xspace}
\providecommand{\sEL}{\ensuremath{\tilde{\mathrm{E}}_\mathrm{L}}\xspace}
\providecommand{\sER}{\ensuremath{\tilde{\mathrm{E}}_\mathrm{R}}\xspace}
\providecommand{\sFer}{\ensuremath{\tilde{\mathrm{f}}}\xspace}
\providecommand{\sQua}{\ensuremath{\tilde{\mathrm{q}}}\xspace}
\providecommand{\sUp}{\ensuremath{\tilde{\mathrm{u}}}\xspace}
\providecommand{\suL}{\ensuremath{\tilde{\mathrm{u}}_\mathrm{L}}\xspace}
\providecommand{\suR}{\ensuremath{\tilde{\mathrm{u}}_\mathrm{R}}\xspace}
\providecommand{\sDw}{\ensuremath{\tilde{\mathrm{d}}}\xspace}
\providecommand{\sdL}{\ensuremath{\tilde{\mathrm{d}}_\mathrm{L}}\xspace}
\providecommand{\sdR}{\ensuremath{\tilde{\mathrm{d}}_\mathrm{R}}\xspace}
\providecommand{\sTop}{\ensuremath{\tilde{\mathrm{t}}}\xspace}
\providecommand{\stL}{\ensuremath{\tilde{\mathrm{t}}_\mathrm{L}}\xspace}
\providecommand{\stR}{\ensuremath{\tilde{\mathrm{t}}_\mathrm{R}}\xspace}
\providecommand{\stone}{\ensuremath{\tilde{\mathrm{t}}_1}\xspace}
\providecommand{\sttwo}{\ensuremath{\tilde{\mathrm{t}}_2}\xspace}
\providecommand{\sBot}{\ensuremath{\tilde{\mathrm{b}}}\xspace}
\providecommand{\sbL}{\ensuremath{\tilde{\mathrm{b}}_\mathrm{L}}\xspace}
\providecommand{\sbR}{\ensuremath{\tilde{\mathrm{b}}_\mathrm{R}}\xspace}
\providecommand{\sbone}{\ensuremath{\tilde{\mathrm{b}}_1}\xspace}
\providecommand{\sbtwo}{\ensuremath{\tilde{\mathrm{b}}_2}\xspace}
\providecommand{\sLep}{\ensuremath{\tilde{\mathrm{l}}}\xspace}
\providecommand{\sLepC}{\ensuremath{\tilde{\mathrm{l}}^\mathrm{C}}\xspace}
\providecommand{\sEl}{\ensuremath{\tilde{\mathrm{e}}}\xspace}
\providecommand{\sElC}{\ensuremath{\tilde{\mathrm{e}}^\mathrm{C}}\xspace}
\providecommand{\seL}{\ensuremath{\tilde{\mathrm{e}}_\mathrm{L}}\xspace}
\providecommand{\seR}{\ensuremath{\tilde{\mathrm{e}}_\mathrm{R}}\xspace}
\providecommand{\snL}{\ensuremath{\tilde{\nu}_L}\xspace}
\providecommand{\sMu}{\ensuremath{\tilde{\mu}}\xspace}
\providecommand{\sNu}{\ensuremath{\tilde{\nu}}\xspace}
\providecommand{\sTau}{\ensuremath{\tilde{\tau}}\xspace}
\providecommand{\Glu}{\ensuremath{\mathrm{g}}\xspace}
\providecommand{\sGlu}{\ensuremath{\tilde{\mathrm{g}}}\xspace}
\providecommand{\Wpm}{\ensuremath{\mathrm{W}^{\pm}}\xspace}
\providecommand{\sWpm}{\ensuremath{\tilde{\mathrm{W}}^{\pm}}\xspace}
\providecommand{\Wz}{\ensuremath{\mathrm{W}^{0}}\xspace}
\providecommand{\sWz}{\ensuremath{\tilde{\mathrm{W}}^{0}}\xspace}
\providecommand{\sWino}{\ensuremath{\tilde{\mathrm{W}}}\xspace}
\providecommand{\Bz}{\ensuremath{\mathrm{B}^{0}}\xspace}
\providecommand{\sBz}{\ensuremath{\tilde{\mathrm{B}}^{0}}\xspace}
\providecommand{\sBino}{\ensuremath{\tilde{\mathrm{B}}}\xspace}
\providecommand{\Zz}{\ensuremath{\mathrm{Z}^{0}}\xspace}
\providecommand{\sZino}{\ensuremath{\tilde{\mathrm{Z}}^{0}}\xspace}
\providecommand{\sGam}{\ensuremath{\tilde{\gamma}}\xspace}
\providecommand{\chiz}{\ensuremath{\tilde{\chi}^{0}}\xspace}
\providecommand{\chip}{\ensuremath{\tilde{\chi}^{+}}\xspace}
\providecommand{\chim}{\ensuremath{\tilde{\chi}^{-}}\xspace}
\providecommand{\chipm}{\ensuremath{\tilde{\chi}^{\pm}}\xspace}
\providecommand{\Hone}{\ensuremath{\mathrm{H}_\mathrm{d}}\xspace}
\providecommand{\sHone}{\ensuremath{\tilde{\mathrm{H}}_\mathrm{d}}\xspace}
\providecommand{\Htwo}{\ensuremath{\mathrm{H}_\mathrm{u}}\xspace}
\providecommand{\sHtwo}{\ensuremath{\tilde{\mathrm{H}}_\mathrm{u}}\xspace}
\providecommand{\sHig}{\ensuremath{\tilde{\mathrm{H}}}\xspace}
\providecommand{\sHa}{\ensuremath{\tilde{\mathrm{H}}_\mathrm{a}}\xspace}
\providecommand{\sHb}{\ensuremath{\tilde{\mathrm{H}}_\mathrm{b}}\xspace}
\providecommand{\sHpm}{\ensuremath{\tilde{\mathrm{H}}^{\pm}}\xspace}
\providecommand{\hz}{\ensuremath{\mathrm{h}^{0}}\xspace}
\providecommand{\Hz}{\ensuremath{\mathrm{H}^{0}}\xspace}
\providecommand{\Az}{\ensuremath{\mathrm{A}^{0}}\xspace}
\providecommand{\Hpm}{\ensuremath{\mathrm{H}^{\pm}}\xspace}
\providecommand{\sGra}{\ensuremath{\tilde{\mathrm{G}}}\xspace}
\providecommand{\mtil}{\ensuremath{\tilde{m}}\xspace}
\providecommand{\rpv}{\ensuremath{\rlap{\kern.2em/}R}\xspace}
\providecommand{\LLE}{\ensuremath{LL\bar{E}}\xspace}
\providecommand{\LQD}{\ensuremath{LQ\bar{D}}\xspace}
\providecommand{\UDD}{\ensuremath{\overline{UDD}}\xspace}
\providecommand{\Lam}{\ensuremath{\lambda}\xspace}
\providecommand{\Lamp}{\ensuremath{\lambda'}\xspace}
\providecommand{\Lampp}{\ensuremath{\lambda''}\xspace}
\providecommand{\spinbd}[2]{\ensuremath{\bar{#1}_{\dot{#2}}}\xspace}

\providecommand{\MD}{\ensuremath{{M_\mathrm{D}}}\xspace}
\providecommand{\Mpl}{\ensuremath{{M_\mathrm{Pl}}}\xspace}
\providecommand{\Rinv} {\ensuremath{{R}^{-1}}\xspace} 
\newcommand{\checkit}[1]{}
\newcommand{\arc}[1]{{\it Note to ARC: #1 }}

\newcommand\brabar{\raisebox{-4.0pt}{\scalebox{.2}{\textbf{(}}}\raisebox{-4.0pt}{{\_}}\raisebox{-4.0pt}{\scalebox{.2}{\textbf{)}}}}

\newcommand{\upsi}  {\ensuremath{\Upsilon\mathrm{(1S)}\,}}
\newcommand{\upsii} {\ensuremath{\Upsilon\mathrm{(2S)}\,}}
\newcommand{\upsiii}{\ensuremath{\Upsilon\mathrm{(3S)}\,}}
\newcommand{\upsn}  {\ensuremath{\Upsilon\mathrm{(nS)}\,}}
\newcommand{\Jpsi}{\ensuremath{{\mathrm{J/\psi~}}}} 
\newcommand{\sqrts} {\sqrt{s}}
\newcommand{\mumu}{\mu^+\mu^-}
\newcommand{\inb}{\ensuremath{\mathrm{nb^{-1}}}\,} 
\newcommand{\ipb}{\ensuremath{\mathrm{pb^{-1}}}\,} 
\newcommand{\eff}{\varepsilon}
\newcommand{\acc}{{\cal A}}
\newcommand{\ups}{\ensuremath{\Upsilon(\mathrm{nS})}}

\cmsNoteHeader{BPH-10-003}
\title{Upsilon production cross section in pp collisions at $\sqrts=7$~TeV}
\address[cern]{CERN}
\author[cern]{CMS}

\date{\today}

\abstract{
  The $\Upsilon$ production cross section in proton-proton collisions at $\sqrt{s} = 7$\,TeV
is measured using a data sample collected with the CMS detector at the LHC,
  corresponding to an integrated luminosity of $3.1\pm 0.3\,\text{pb}^{-1}$.
  Integrated over the rapidity range $|y|<2$, we find
  the product of the $\Upsilon(1S)$ production cross section and
  branching fraction to dimuons to be $\sigma(\mathrm{pp} \rightarrow \Upsilon {
    (1S)} X ) \cdot {\cal B} (\Upsilon(1S) \rightarrow \mu^+\mu^-) =
  7.37 \pm 0.13^{+0.61}_{-0.42}\pm 0.81$\,nb,
  where the first   uncertainty is statistical, the second is systematic, and the third
  is associated with the estimation of the integrated luminosity of the data sample.
  This cross section is obtained assuming unpolarized $\Upsilon(1S)$ production.
  If the $\Upsilon(1S)$ production polarization is fully transverse or fully longitudinal the cross section changes by about 20\,\%.
  We also report the measurement of the $\Upsilon(1S)$, $\Upsilon(2S)$, and $\Upsilon(3S)$ differential cross sections as a function of transverse momentum and rapidity.
}

\hypersetup{%
pdfauthor={CMS Collaboration},%
pdftitle={Measurement of the Inclusive Upsilon production cross section in \Pp\Pp collisions at $\sqrts=7$\TeV},%
pdfsubject={CMS},%
pdfkeywords={CMS, physics, quarkonia}}

\maketitle 

\section{Introduction}
\label{sec:introduction}

The hadroproduction of quarkonia is not understood.
None of the existing theories successfully reproduces both the differential cross section and the polarization measurements of the $\JPsi$ or $\PgU$ states~\cite{yellow}.
It is expected that studying quarkonium hadroproduction at higher center-of-mass energies and over a wider rapidity range will facilitate significant improvements in our understanding.
Measurements of the $\PgU$ resonances are particularly important
since the theoretical calculations are more robust
than for the charmonium family, due to the heavy bottom 
quark and the absence of $b$-hadron feed-down.
Measurements of quarkonium hadroproduction cross sections and production polarizations made at the Large Hadron Collider (LHC) will allow important tests of several alternative theoretical approaches. These include non-relativistic QCD (NRQCD) factorization~\cite{bib-nrqcd}, where quarkonium production includes color-octet components, and calculations made in the color-singlet model including next-to-leading order (NLO) corrections~\cite{Artoisenet:2008fc} which reproduce the differential cross sections measured at the Tevatron experiments~\cite{bib-cdfups,bib-d0ups} without requiring a significant color-octet contribution.

This paper presents the first measurement of the $\PgUa$, $\PgUb$,
and $\PgUc$ production cross sections in proton-proton collisions at $\sqrts=7\TeV$,
using data recorded by the Compact Muon
Solenoid (CMS) experiment between April and September 2010.
In these measurements the signal efficiencies are determined with data. Consequently,
Monte Carlo simulation is used only in the evaluation of the geometric and kinematic acceptances.
The document is organized as follows.
Section~\ref{sec:detector} contains a short description of the CMS detector.
Section~\ref{selection} presents the data collection, the online and offline event selections, the $\PgU$~reconstruction, and the Monte Carlo simulation.
The detector acceptance and efficiencies to reconstruct $\PgU$~events in CMS are discussed in Sections~\ref{sec:acceptance} and~\ref{efficiency}, respectively.
In Section~\ref{crosssection} the fitting technique employed to extract the cross section is presented.
The evaluation of systematic uncertainties on the 
measurements is described in Section~\ref{sec:systematics}.
Section~\ref{summary} presents the $\PgU(nS)$ cross section results and comparisons to existing measurements at lower collision energies~\cite{bib-cdfups,bib-d0ups} and to the \PYTHIA~\cite{bib-PYTHIA} event generator.

\section{The CMS detector}
\label{sec:detector}

The central feature of the CMS apparatus is a superconducting solenoid, of 6\unit{m} inner diameter, providing a field of 3.8\unit{T}.
Inside the solenoid in order of increasing distance from the interaction point are the silicon pixel and strip tracker, the crystal electromagnetic calorimeter, and the brass/scintillator hadron calorimeter.
Muons are detected by three types of gas-ionization detectors embedded in the steel return yoke:
drift tubes (DT), cathode strip chambers (CSC), and resistive plate chambers (RPC).
The muon measurement covers the pseudorapidity range $|\eta|< 2.4$, where $\eta=-{\ln}[{\text{tan}}(\theta/2)]$ and the polar angle $\theta$ is measured from the $z$-axis, which points along the counterclockwise beam direction.
The silicon tracker is composed of pixel detectors (three barrel
layers and two forward disks on either side of the detector, made of
66~million $100\cdot150\micron^2$ pixels), followed by microstrip detectors
(ten barrel layers plus three inner disks and nine forward disks on either side of the
detector, with 10 million strips of pitch between 80 and 184\micron).
Due to the strong magnetic field and the fine granularity of the silicon tracker, the transverse momentum,
\pt, of the muons matched to silicon tracks is measured with a
resolution of about 1\% for a typical muon in this analysis.
The silicon tracker also provides the primary vertex position with $\sim$\,20\micron accuracy.
The first level (L1) of the CMS trigger system, composed of custom
hardware processors, uses information from the calorimeters and muon
detectors to select the most interesting events.  The high-level trigger (HLT)
further decreases the event rate before data storage.
A more detailed description of the CMS detector can be found elsewhere~\cite{JINST}.

\section{Data sample and event reconstruction}
\label{selection}

\subsection{Event selection}

The data sample used in this analysis was recorded by the CMS detector
in \Pp\Pp collisions at a center-of-mass energy of 7\TeV. The sample
corresponds to a total integrated luminosity of $3.1 \pm 0.3\pbinv$~\cite{bib-lumi-pas}.
The maximum instantaneous luminosity was $10^{31}\percms$ and event pileup was negligible.
Data are included in the analysis  if the silicon tracker, the muon detectors, and the
trigger were performing well and the luminosity measurement was available.

The trigger requires the detection of two muons at the hardware level,
without any further selection at the HLT. The coincidence of two muon signals
without an explicit \pt requirement is sufficient to maintain
the dimuon trigger without prescaling.
All three muon systems -- DT, CSC, and RPC -- take part in the trigger decision.

Anomalous events arising from beam-gas interactions or beam scraping
in the beam transport system near the interaction point, which produce
a large number of hits in the pixel detector, are removed with offline
software filters~\cite{bib-cms-track}.
A good primary vertex is also required, as defined in Ref.~\cite{bib-cms-track}.
The detector systems are aligned and calibrated using LHC collision
data and cosmic-ray muons~\cite{bib-trackeralignment}.

\subsection{Monte Carlo simulation}
Simulated events are used to tune the selection criteria and to
compare with data. Upsilon events are produced using
\PYTHIA~6.412~\cite{bib-PYTHIA},
which generates events based on the
leading-order color-singlet and octet mechanisms,
with  NRQCD matrix elements tuned
by comparing calculations with the CDF data~\cite{kramer} and
applying the normalization and wavefunctions as recommended in Ref.~\cite{bib-marianne}.
The generation of $\PgUb$ and $\PgUc$ states has been included by changing the relevant particle masses and branching ratios.
The simulation includes the generation of $\chi_b$ states.
Final-state radiation (FSR) is implemented using \PHOTOS~\cite{bib-photos1,bib-photos2}.
The response of the CMS detector is simulated with a \GEANT4-based~\cite{bib-GEANT4}
Monte Carlo (MC) program. The simulated events are processed with the same
reconstruction algorithms as used for data.

\subsection{Offline muon reconstruction}

In this analysis, a muon is defined as a track reconstructed in the
silicon tracker and associated with a compatible signal in the muon
detectors.
Tracks are reconstructed using a Kalman filter technique
which starts from hits in the pixel system and extrapolates outward to
the silicon strip tracker.
The tracks found in the silicon tracker are  propagated to the muon system and
required to be matched to at least one muon segment in one muon station.
Further details may be found in Ref.~\cite{bib-muonreco}.

Quality criteria are applied to the tracks to reject 
muons from kaon and pion decays. The tracks are
required to have at least twelve hits in the silicon tracker, at least
one of which must be in the pixel detector, and a track-fit $\chi^2$ per degree of
freedom smaller than five. In addition the tracks are required to emanate from a cylinder of radius
2\mm and length 50\cm centered on the $\Pp\Pp$ interaction region
and parallel to the beam line.
The muons are required to satisfy:
\ifthenelse{\boolean{cms@external}}{
\begin{align}
  \label{eq:selection}
      \pt^\mu > 3.5\GeVc& \; \text{\;if\;}  \; |\eta^\mu|<1.6, \; \text{or} \\
      \pt^\mu > 2.5\GeVc& \; \text{\;if\;}  \; 1.6<|\eta^\mu|<2.4 \,.
    \end{align}
}{
\begin{equation}
  \label{eq:selection}
     \pt^\mu > 3.5\GeVc \quad \text{if} \quad  |\eta^\mu|<1.6 \;, \quad \text{or} \quad \pt^\mu > 2.5\GeVc \quad {\rm  if} \quad   1.6<|\eta^\mu|<2.4 \;.
\end{equation}
}
These kinematic criteria are chosen to ensure that the
trigger and muon reconstruction efficiencies are high and not rapidly
changing within the acceptance window of the analysis.

The momentum measurement of charged tracks in the CMS detector is
affected by systematic uncertainties caused by imperfect knowledge of
the magnetic field, the amount of material, and
sub-detector misalignments, as well as by biases in the algorithms
which fit the track trajectory.
A mismeasurement of the track momentum results in a shift and broadening of the reconstructed peaks of dimuon resonances.
An improved understanding of the CMS magnetic field, detector alignment, and material budget was obtained
from cosmic-ray muon and LHC collision data~\cite{bib-trackeralignment, bib-magneticfield, bib-material},
and the residual effects are determined by studying
the dependence of the reconstructed $\JPsi$ dimuon invariant-mass
distribution on the muon kinematics~\cite{bib-trackermomentum}.
The transverse momentum corrected for the residual scale distortion is parameterized as
$\pt  =  (1+ a_1 + a_2 \eta^2) \cdot \pt^{\prime}$,
where $\pt^{\prime}$ is the measured muon transverse momentum,
 $ a_1 = (3.8 \pm 1.9) \cdot 10^{-4}$, and
$a_2 = (3.0  \pm 0.7) \cdot 10^{-4}$.
The coefficients for terms linear in $\eta$ and quadratic in $p_T^{\prime}$ and $p_T^{\prime}\cdot\eta$ are consistent with zero and are not included.

\subsection{\texorpdfstring{$\mathbf {\PgU}$~event selection}{Upsilon event selection}}
\label{sec:ups_selection}

To identify events containing an $\PgU$~decay, muons with opposite
charges are paired, and the invariant mass of the muon pair is required to be between 8 and 14\GeVcc.
The longitudinal separation between the two muons
at their points of closest approach to the beam axis is required to be less than 2\cm.
The two muon helices are fit with a common vertex constraint, and
events are retained if the fit $\chi^2$ probability is larger than 0.1\%.
The dimuon candidate is confirmed offline to have passed the trigger requirements.
If multiple dimuon candidates are found in the same event, the candidate with the best
vertex quality is retained;
the fraction of signal candidates rejected by this requirement is about 0.2\%.
Finally, the rapidity, $y$, of the $\PgU$ candidates is required to satisfy $|y|<2$ because the acceptance
diminishes rapidly at larger rapidity.
The rapidity is defined as $y=\frac{1}{2}\,\ln\left(\frac{E+p_\parallel}{E-p_\parallel}\right)$,
where $E$ is the energy and $p_\parallel$ the momentum parallel to the beam axis of the muon pair.

\begin{figure}[htp]
  \centering
  \includegraphics[angle=0,width=0.49\textwidth]{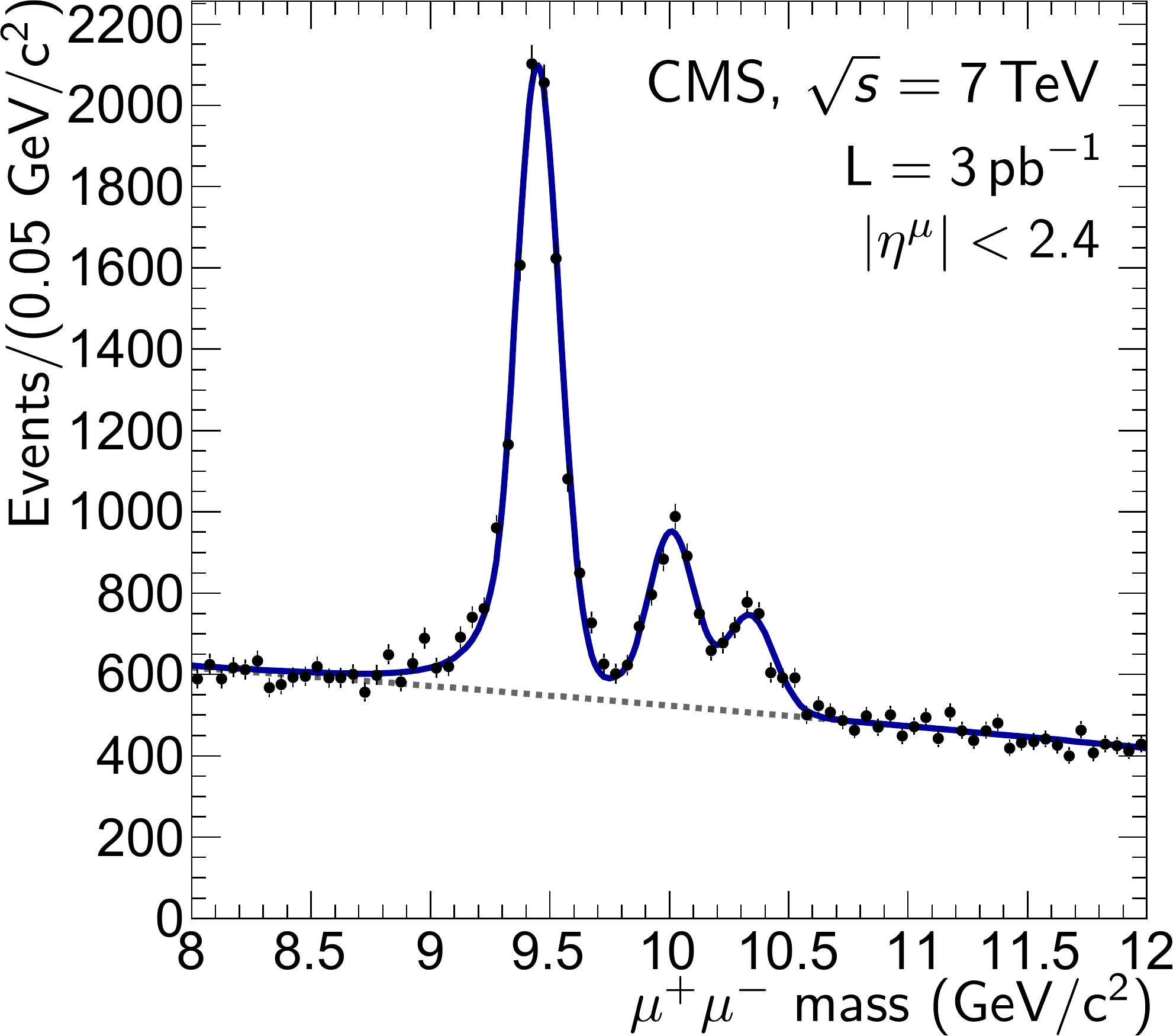}
  \includegraphics[angle=0,width=0.49\textwidth]{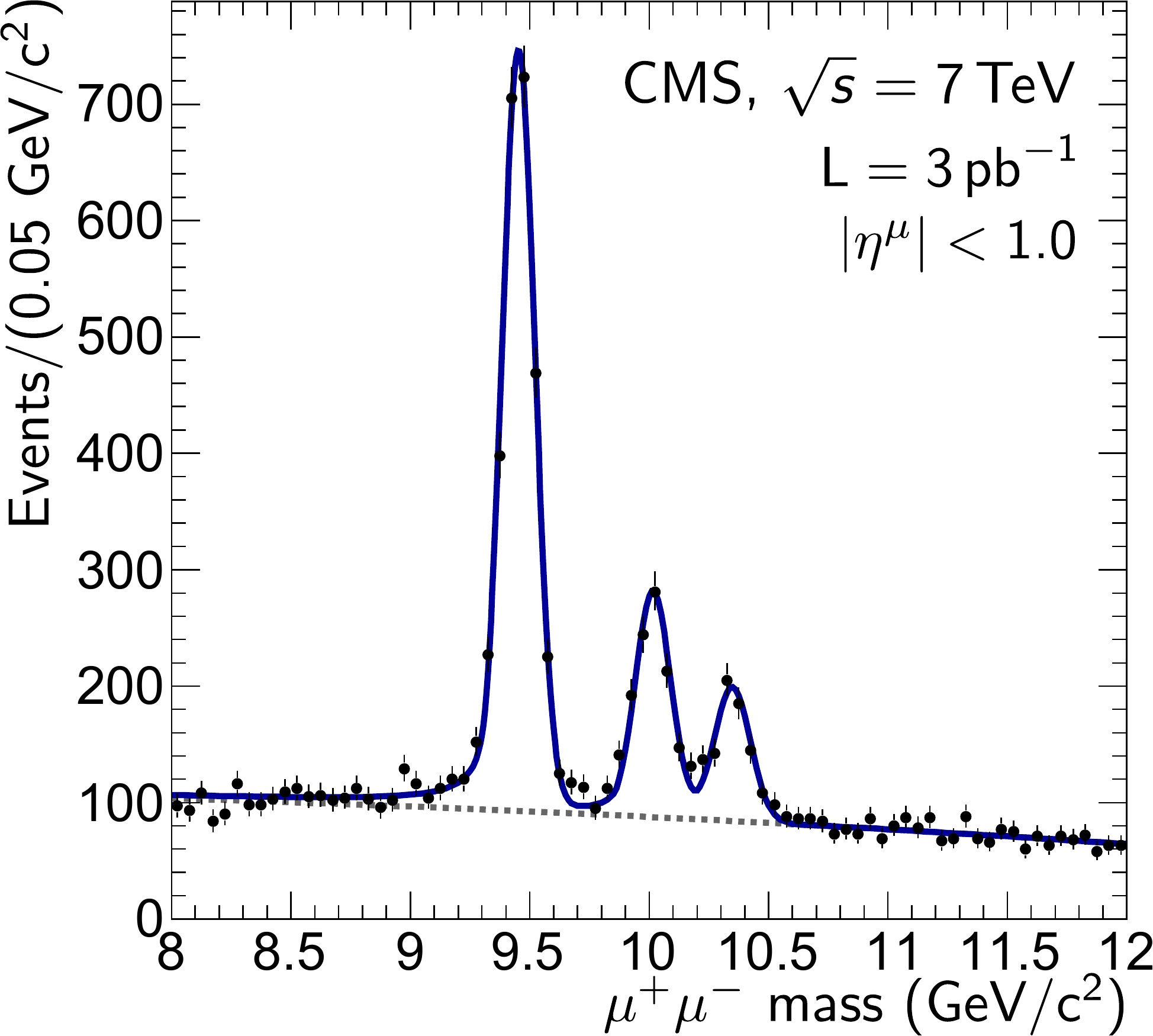}
  \caption{The dimuon invariant-mass distribution in the vicinity of
    the $\PgU(nS)$ resonances for the full rapidity covered by the analysis (left) and for the subset of events where the pseudorapidity of each muon satisfies $|\eta^{\mu}|<1$ (right).
    The solid line shows the result of a fit to the invariant-mass distribution before accounting for acceptance and efficiency, with the dashed line  denoting the background component.
    Details of the fit are described in Section~\ref{crosssection}.}
  \label{fig:massFit-raw}
\end{figure}

The dimuon invariant-mass spectrum in the $\PgU(nS)$ region for the dimuon transverse momentum interval
$\pt < 30\GeVc$ is shown in Fig.~\ref{fig:massFit-raw} for the pseudorapidity intervals $|\eta^\mu|<2.4$ (left) and $|\eta^\mu|<1.0$ (right).
We obtain a $\PgUa$ mass resolution of $96\pm2\MeVcc$ including muons from the entire pseudorapidity range, and $69\pm2\MeVcc$ when both muons satisfy $|\eta^{\mu}|<1$.
The observed resolutions, determined as a parameter of the fit function described in Section~\ref{crosssection}, are in good agreement with the predictions from MC simulation.

\section{Acceptance}
\label{sec:acceptance}

The $\PgU \to \mu^+ \mu^-$ acceptance of the CMS detector is defined as the
product of two terms. The first is the fraction of upsilons 
of given \pt and $y$ such that each of the two muons satisfies
Eq.~(\ref{eq:selection}).  The second is the probability that each muon can be reconstructed in the tracker using the CMS software, in the absence of further event activity in the detector and without quality cuts imposed.
Both components are evaluated by simulation and parametrized as a function
of the \pt and rapidity of the $\PgU$.

The acceptance is calculated from the ratio
\begin{equation}
  {\mathcal A^\PgU}\left(\pt, y \right) =
  \frac{N^\PgU_{\rm rec}\left(\pt,  y \right)}
  {N^\PgU_{\rm gen}\left( \pt,  y \right)}\;,
\label{eqn:acc}
\end{equation}
where $N^\PgU_{\rm gen}\left( \pt,  y \right)$ is the
number of upsilons generated in a $( \pt, y)$ bin, while
$N^\PgU_{\rm rec}\left( \pt,  y \right)$ is the number
reconstructed in the same $( \pt, y)$ region but
now using the reconstructed, rather than generated, variables.
In addition, the numerator requires that the two
muons reconstructed in the silicon tracker
satisfy Eq.~(\ref{eq:selection}).

The acceptance is evaluated with a signal MC sample in which the $\PgU$ decay to two muons is generated with the
\EVTGEN~\cite{bib-evtgen} package including the effects of final-state radiation.
There are no particles in the event besides the
$\PgU$, its daughter muons, and final-state radiation.
The upsilons are generated uniformly in \pt and rapidity.
This sample is then fully simulated and reconstructed with the
CMS detector simulation software to assess the effects of
multiple scattering and finite resolution of the detector.
The acceptance is calculated for two-dimensional (2-D) bins of size ($1\GeVc \cdot 0.1$)
in the reconstructed \pt and $y$ of the $\PgU$
and used in candidate-by-candidate yield corrections.

\begin{figure}[htp]
  \centering
  \includegraphics[angle=90,width=0.47\textwidth]{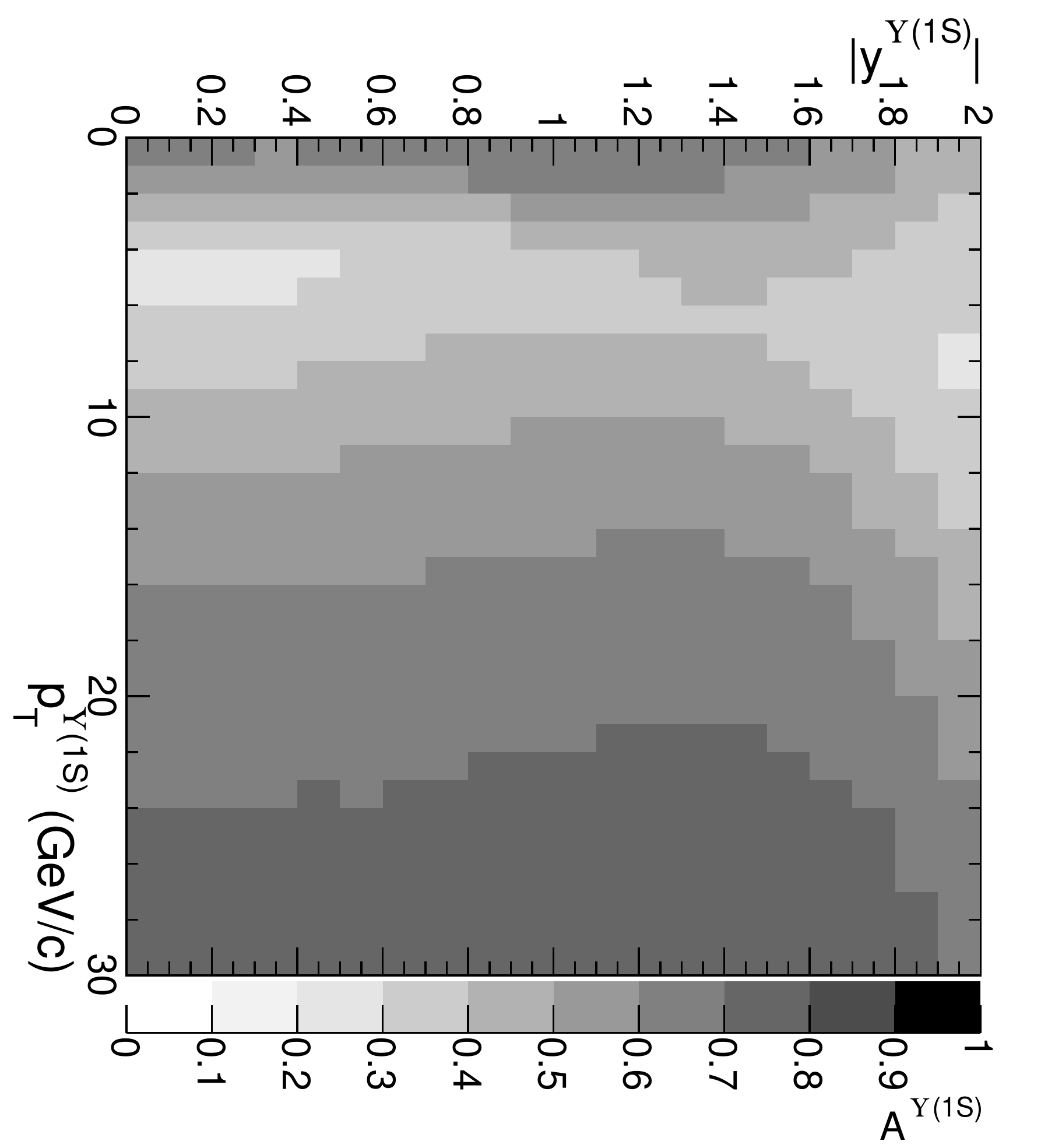}
  \includegraphics[angle=90,width=0.47\textwidth]{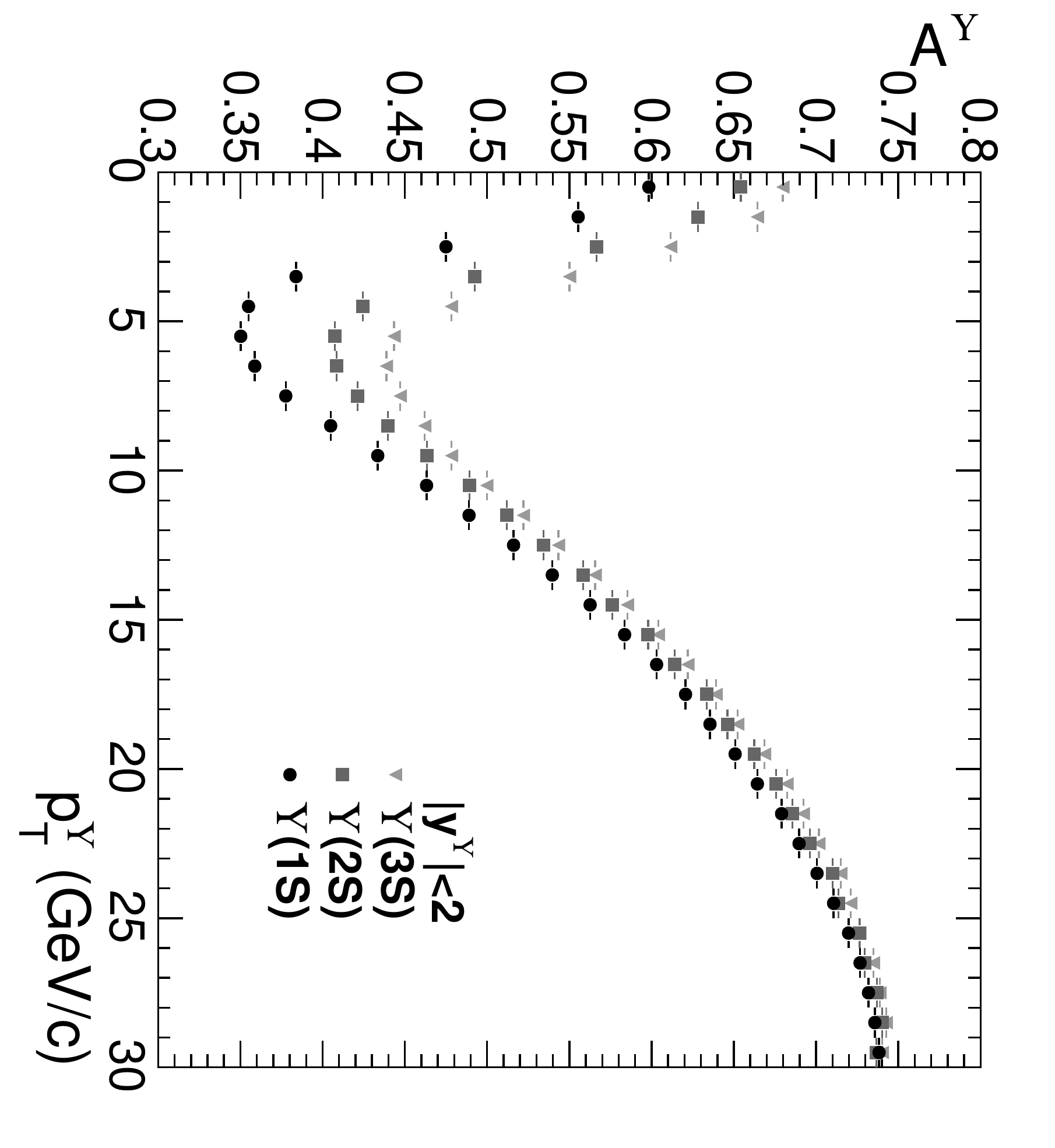}
  \caption{(Left) Unpolarized $\PgUa$ acceptance as a function of \pt and $y$;
    (Right) the unpolarized $\PgUa$, $\PgUb$, and $\PgUc$ acceptances integrated over rapidity as a function of \pt.
    \label{fig:acceptUnpol}}
\end{figure}

The 2-D acceptance map for unpolarized $\PgUa$ is shown in the left plot of Fig.~\ref{fig:acceptUnpol}.
The acceptance varies with the resonance mass.
This is shown in the right plot of Fig.~\ref{fig:acceptUnpol}, which displays the acceptance integrated over the rapidity range as a function of \pt for each upsilon resonance. 
The transverse-momentum threshold for muon detection, especially in the forward region, is small compared to the upsilon mass.
 Therefore, when the $\PgU$ decays at rest, both muons are likely to reach the muon detector. When the $\PgU$ has a modest boost, the probability is greater that one muon will be below the muon detection threshold and the acceptance decreases until the $\PgU$ transverse momentum reaches about 5\GeVc, after which the acceptance rises slowly.
The production polarization of the $\PgU$ strongly influences the
muon angular distributions and is expected to change as a function of
\pt.  In order to account for this, the acceptance is calculated for
five extreme polarization scenarios~\cite{bib-faccioli}: unpolarized and polarized
longitudinally and transversely with respect to two different
reference frames. The first is the helicity frame (HX), defined by the
flight direction of the $\PgU$ in the center-of-mass system of the
colliding beams. The second is the Collins-Soper (CS)
frame~\cite{bib-CollinsSoper}, defined by the bisection of the
incoming proton directions in the $\PgU$ rest frame.

\section{Efficiency}
\label{efficiency}

\begin{table}[t]
  \centering
  \renewcommand{\arraystretch}{1.3}
  \caption{Single-muon identification efficiencies, in percent, measured from $\JPsi$ data with T\&P.
    The statistical uncertainties in the least significant digits are given in parentheses;
    uncertainties less than 0.05 are denoted by 0. For asymmetric uncertainties the positive uncertainty is reported first.
    \newline}
  \begin{tabular}{r@{--}l r@{}l r@{}l r@{}l r@{}l r@{}l r@{}l }\hline
\multicolumn{2}{c}{$p_T^\mu$} & \multicolumn{12}{c}{$|\eta^\mu|$} \\ \cline{3-14}
\multicolumn{2}{c}{$(\!\GeVc)$}  & \multicolumn{2}{c}{0.0--0.4} &  \multicolumn{2}{c}{0.4--0.8} &  \multicolumn{2}{c}{0.8--1.2} &  \multicolumn{2}{c}{1.2--1.6} &  \multicolumn{2}{c}{1.6--2.0}  &  \multicolumn{2}{c}{2.0--2.4} \\ \hline
2.5&3.0 &      &        &     &        &      &        &      &        & $100$&$(0,4)$ & $ 94$&$(6  )$ \\
3.0&3.5 &      &        &     &        &      &        &      &        & $ 95$&$(3  )$ & $100$&$(0,4)$ \\
3.5&4.0 & $83 $&$(2)  $ &$ 89$&$(2)  $ & $ 88$&$(2  )$ & $ 96$&$(3)  $ & $100$&$(0,2)$ & $100$&$(0,4)$ \\
4.0&4.5 & $92 $&$(2)  $ &$ 95$&$(2)  $ & $ 99$&$(1,3)$ & $ 98$&$(2,3)$ & $100$&$(0,2)$ & $100$&$(0,6)$ \\
4.5&5.0 & $99 $&$(1,2)$ &$ 99$&$(1,3)$ & $ 95$&$(3  )$ & $ 96$&$(3  )$ & $100$&$(0,2)$ & $100$&$(0,4)$ \\
5.0&6.0 & $98 $&$(2)  $ &$100$&$(0,1)$ & $100$&$(0,2)$ & $100$&$(0,1)$ & $ 97$&$(3  )$ & $100$&$(0,5)$ \\
6.0&8.0 & $100$&$(0,2)$ &$100$&$(0,1)$ & $100$&$(0,1)$ & $100$&$(0,2)$ & $100$&$(0,2)$ & $ 94$&$(6,7)$ \\
8.0&50.0& $100$&$(0,2)$ & $97$&$(3)  $ & $100$&$(0,3)$ & $ 97$&$(3,4)$ & $100$&$(0,3)$ & $ 98$&$(2,9)$ \\
 \hline
\end{tabular}

  \label{tab:TMIDEffs}
  \vspace{5mm}
  \caption{Single-muon trigger efficiencies, in percent, measured from $\JPsi$ data with T\&P.
    The statistical uncertainties in the least significant digits are given in  parentheses;
    uncertainties less than 0.05 are denoted by 0. For asymmetric uncertainties the positive uncertainty is reported first.
    \newline}
  \label{tab:TrigEffs}
  \begin{tabular}{r@{--}l r@{}l r@{}l r@{}l r@{}l r@{}l r@{}l }\hline
\multicolumn{2}{c}{$p_T^\mu$} & \multicolumn{12}{c}{$|\eta^\mu|$} \\ \cline{3-14}
\multicolumn{2}{c}{$(\!\GeVc)$}  & \multicolumn{2}{c}{0.0--0.4} &  \multicolumn{2}{c}{0.4--0.8} &  \multicolumn{2}{c}{0.8--1.2} &  \multicolumn{2}{c}{1.2--1.6} &  \multicolumn{2}{c}{1.6--2.0}  &  \multicolumn{2}{c}{2.0--2.4} \\ \hline
2.5&3.0 &     &      &     &     &      &      &      &          & $93$&$(1)$ & $92$&$(2)$ \\
3.0&3.5 &     &      &     &     &      &      &      &          & $94$&$(1)$ & $93$&$(1)$ \\
3.5&4.0 & $69$&$(1)$ & $81$&$(1)$ & $78$&$(1)$ & $ 98$&$(1)$     & $94$&$(1)$ & $97$&$(1)$ \\
4.0&4.5 & $79$&$(1)$ & $91$&$(1)$ & $86$&$(1)$ & $ 98$&$(1)$     & $92$&$(1)$ & $96$&$(1)$ \\
4.5&5.0 & $85$&$(1)$ & $95$&$(1)$ & $87$&$(1)$ & $ 97$&$(1)$     & $96$&$(1)$ & $99$&$(1)$ \\
5.0&6.0 & $90$&$(1)$ & $97$&$(1)$ & $85$&$(1)$ & $ 99$&$(0,1)$   & $95$&$(1)$ & $96$&$(1)$ \\
6.0&8.0 & $92$&$(1)$ & $97$&$(1)$ & $85$&$(1)$ & $100$&$(0)$     & $97$&$(1)$ & $99$&$(1)$ \\
8.0&50.0& $92$&$(1)$ & $97$&$(1)$ & $86$&$(1)$ & $ 99$&$(1)$     & $97$&$(1)$ & $99$&$(2)$ \\
 \hline
\end{tabular}

\end{table}

\begin{figure*}[htp]
  \begin{center}
\includegraphics[angle=0,width=0.9\textwidth]{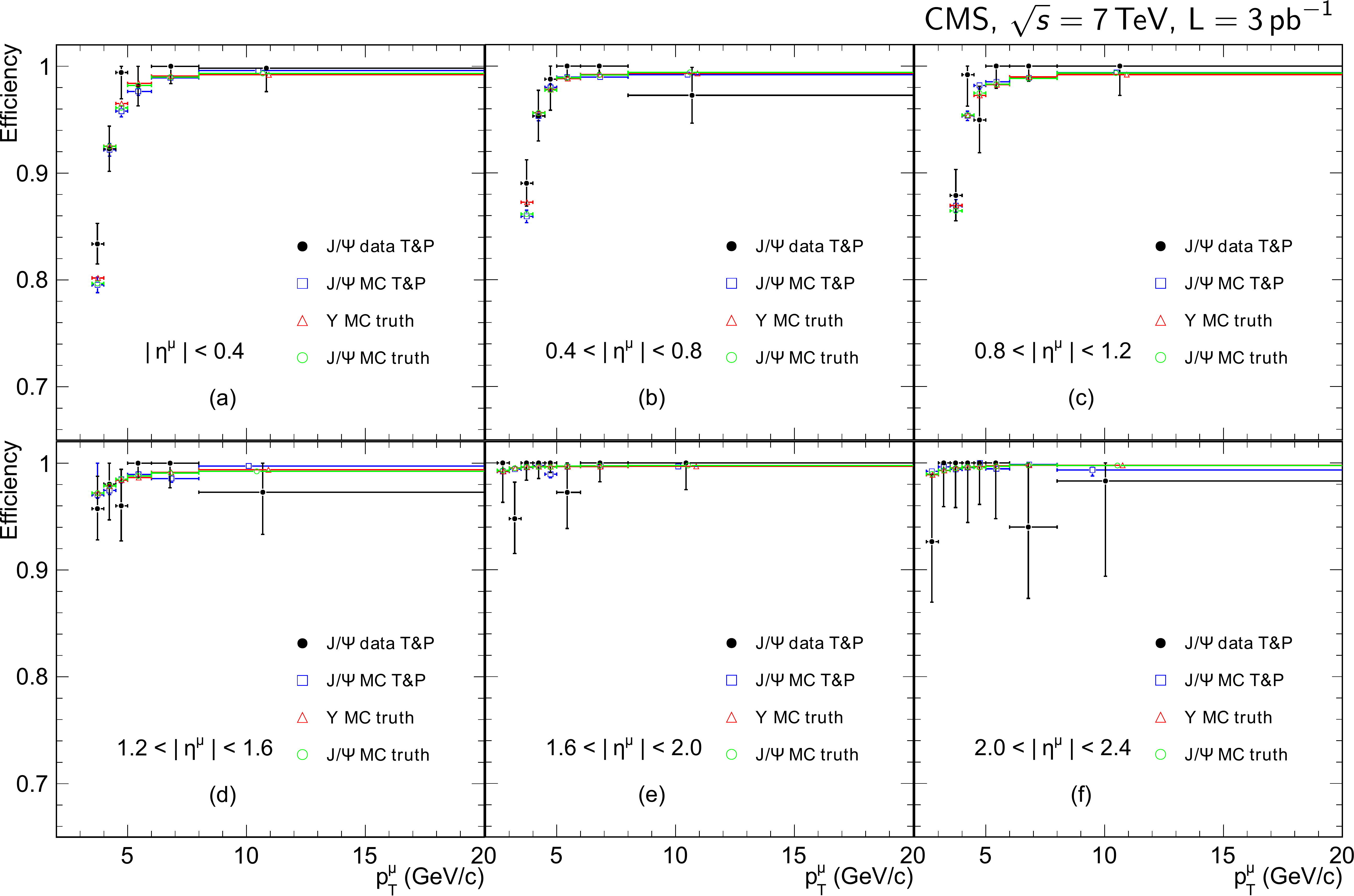}
  \end{center}
  \caption{Single-muon identification efficiencies as a function of $\pt^\mu$ for six $|\eta^\mu|$ regions, measured from data using $\JPsi$ T\&P (closed circles). The efficiencies determined with $\PgU$ MC truth (triangles), $\JPsi$ MC truth (open circles), and $\JPsi$ MC T\&P (squares), used in the evaluation of systematic uncertainties, are also shown.}
  \label{fig:TurnOnMuId}
\end{figure*}

\begin{figure*}[htp]
  \centering
\includegraphics[angle=0,width=0.9\textwidth]{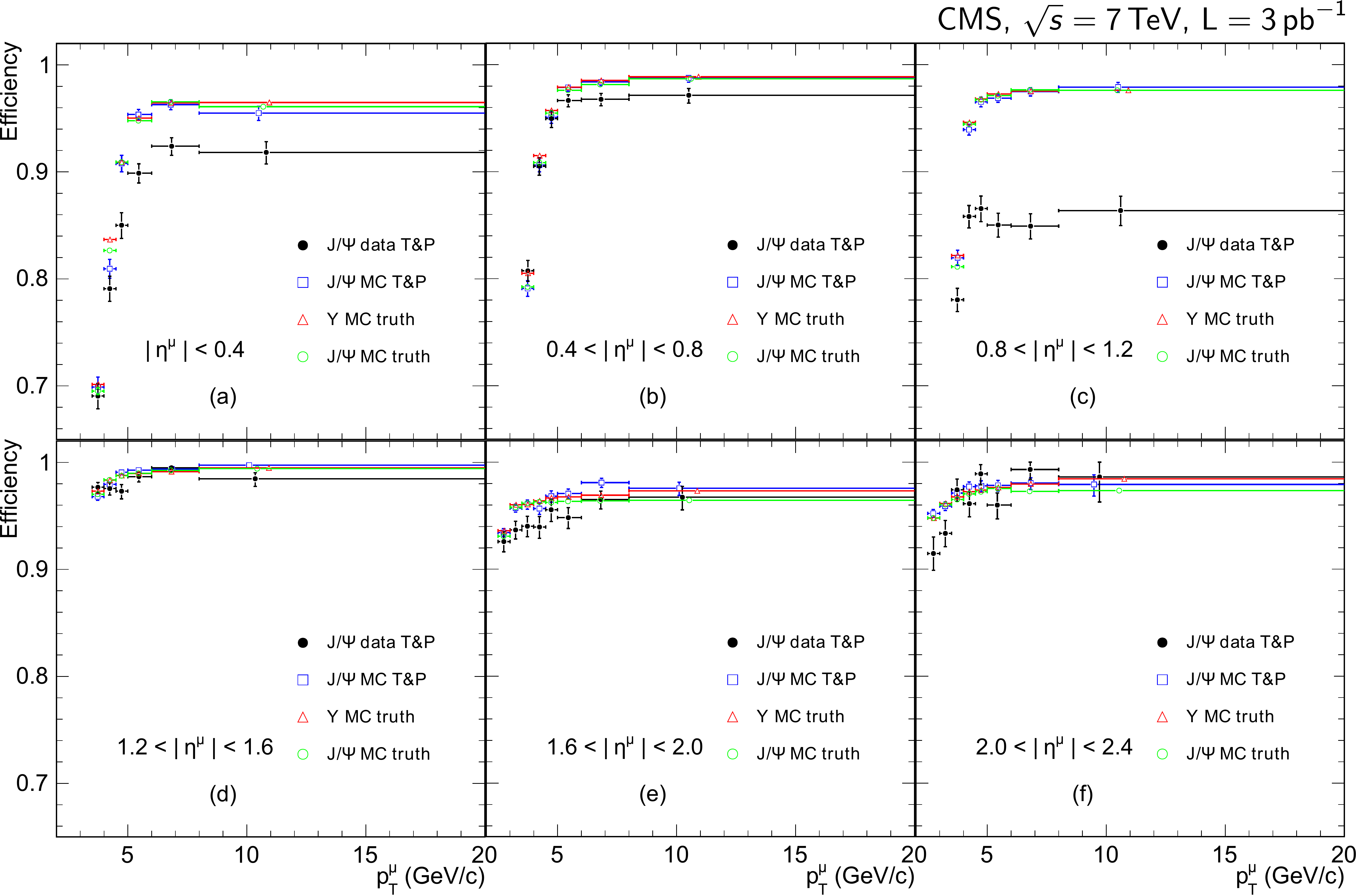}
\caption{Single-muon trigger efficiencies as a function of $\pt^\mu$ for six $|\eta^\mu|$ regions, measured from data using $\JPsi$ T\&P (closed circles). The efficiencies determined with $\PgU$ MC truth (triangles), $\JPsi$ MC truth (open circles), and $\JPsi$ MC T\&P (squares), used in the evaluation of systematic uncertainties, are also shown.}
  \label{fig:TurnOnTrig}
\end{figure*}

We factor the total muon efficiency into three conditional terms,
\ifthenelse{\boolean{cms@external}}{
\begin{align}
  \eff ({\rm total}) &= \eff ({\rm trig | \rm id}) \cdot \eff({\rm id
    | track}) \cdot \eff({\rm track | accepted}) \\
  &\equiv \eff_{\rm trig} \cdot \eff_{\rm id} \cdot \eff_{\rm track}\,.
\end{align}
}{
\begin{equation}
  \eff ({\rm total}) = \eff ({\rm trig | \rm id}) \cdot \eff({\rm id
    | track}) \cdot \eff({\rm track | accepted}) \equiv \eff_{\rm
    trig} \cdot \eff_{\rm id} \cdot \eff_{\rm track}\,.
\end{equation}
}
The tracking efficiency, $\eff_{\rm track}$, combines the efficiency that the accepted track of a muon from
the $\PgU {\rm (nS)}$ decay is reconstructed in the presence of other activity in the silicon tracker,
as determined with a track-embedding technique~\cite{bib-trackingefficiency}, and the efficiency for the track to satisfy quality criteria,
as determined with the tag-and-probe (T\&P) technique~\cite{bib-trackingefficiency} described below. The muon identification
efficiency, $\eff_{\rm id}$, is the probability that the track in the silicon tracker is identified as a muon.
It has been computed as described in Ref.~\cite{bib-muonreco} and is also based on the T\&P method.
The efficiency that an identified muon satisfies the trigger, $\eff_{\rm trig}$, is again measured with the same technique.

The tag and probe technique is a data-based method used in this analysis
to determine the track quality, muon trigger, and muon identification
efficiencies. It utilizes dimuons from $\JPsi$ decays to provide a sample of probe objects.  A
well-identified muon, the tag, is combined with a second object in
the event, the probe, and the invariant mass is computed.  The
tag-probe pairs are divided into two samples, depending on whether the
probe satisfies or not the criteria for the efficiency being evaluated.
The two tag-probe mass distributions contain a
$\JPsi$ peak. The integral of the peak is the number of probes that
satisfy or fail to satisfy the imposed criteria. The efficiency parameter is
extracted from a simultaneous unbinned maximum-likelihood fit to both
mass distributions.

The $\JPsi$ resonance is utilized for T\&P efficiency measurements as it provides a large-yield and statistically-independent dimuon sample~\cite{bib-jpsi-cross-section-BPH-10-002}.
To avoid trigger bias, events containing a tag and probe pair have been collected with triggers that do not impose requirements on the probe from the detector subsystem related to the efficiency measurement.
For the track-quality efficiency measurement, the trigger requires two muons at L1 in the muon system without using the silicon tracker.
For the muon-identification and trigger efficiencies, the trigger requires a muon at the HLT, that is matched to the tag,
paired with a silicon track of opposite sign and the invariant mass of the pair is required to be in the vicinity of the $\JPsi$ mass.

The component of the tracking efficiency measured with the track-embedding technique is well described by a constant value of $(99.64 \pm 0.05)\%$.
The efficiency of the track-quality criteria measured by the T\&P method 
is likewise nearly uniform and has an average value of $(98.66 \pm 0.05)\%$.
Tracks satisfying the quality criteria are the probes for the
muon identification study.
The resulting single-muon identification efficiencies as a function of
$\pt^\mu$ for six $|\eta^\mu|$ regions are
shown in Fig.~\ref{fig:TurnOnMuId} and Table~\ref{tab:TMIDEffs}.
The probes that satisfy the muon
identification criteria are in turn the probes for the
study of the trigger efficiency.
The resulting trigger efficiencies for the same
$\pt^\mu$ and  $|\eta^\mu|$ regions are
shown in Fig.~\ref{fig:TurnOnTrig} and Table~\ref{tab:TrigEffs}.

Figures~\ref{fig:TurnOnMuId} and~\ref{fig:TurnOnTrig} also show single-muon identification and trigger efficiencies, respectively, determined from a high-statistics MC simulation.
The single-muon efficiencies determined with the T\&P technique in the data are found to be consistent, within the uncertainties and  over most of the kinematic range of interest, with the efficiencies obtained from  the $\PgU$ MC simulation utilizing the generator-level particle information (``MC truth''). Two exceptions are the single-muon trigger efficiency for the intervals $|\eta^\mu|<0.4$ and $0.8<|\eta^\mu|<1.2$, where the efficiency is lower in data than in the MC simulation.
In both cases the MC simulation is known to not fully reproduce the detector properties or performance:
gaps in the DT coverage $(|\eta^\mu| < 0.4)$ and suboptimal timing synchronization between the
overlapping CSC and DT subsystems $(0.8<|\eta^\mu| < 1.2)$.
For all cases the data-determined efficiencies are used to obtain the central results.

The $\PgU$ efficiency is estimated from the product of single-muon efficiencies.
Differences between the single and dimuon efficiencies determined from MC truth and those measured with the T\&P technique can arise  from the kinematic distributions of the probes and from bin averaging.
This is evaluated by comparing the single-muon  and dimuon efficiencies as determined using the T\&P method in $\JPsi\to\mu^+\mu^-$ MC events to the efficiencies obtained in the same events utilizing generator-level particle information.
In addition, effects arising from differences in the kinematic distributions between the $\PgU$ and $\JPsi$ decay muons are investigated by comparing the  
efficiencies
determined from $\PgU\to\mu^+\mu^-$ MC events to those from  $\JPsi\to\mu^+\mu^-$ MC events.
In all cases the differences in the efficiency values are not significant, and are 
used only as an estimate of the associated systematic uncertainties.

The efficiency of the vertex $\chi^2$ probability cut is determined using the high-statistics $\JPsi$ data sample, to which the $\PgU$ selection criteria are applied.
The efficiency is extracted from a simultaneous fit to the dimuon mass distribution of the passing and failing candidates. It is found to be $(99.2 \pm 0.1)\%$.
A possible difference between the efficiency of the vertex $\chi^2$ probability cut for the
$\JPsi$ and $\PgU$ is evaluated by applying the same technique to
large MC signal samples of each resonance. No significant
difference in the efficiencies is found.
The efficiency of the remaining selection criteria listed in Section~\ref{selection} is studied in data and MC simulation and is found to be consistent with unity.

\section{Measurement of the cross sections}
\label{crosssection}

The $\PgU(nS)$ differential cross section is determined from the signal yield, $N^{\rm{fit}}$,  obtained directly from a weighted fit to the dimuon invariant-mass spectrum, after correcting for the acceptance (${\cal{A}}$) and the total efficiency ($\eff$), through the equation
\begin{align}
  \label{eqn:xsection}
      \frac{ d^{2}\sigma\left( \Pp\Pp \rightarrow \PgU(nS)X\right) } { d\pt \, dy}
&\cdot {\cal B} \left(\PgU(nS)\rightarrow\mu^+\mu^-\right) \\
& =
      \frac{ N^{\text{fit}}_{\PgU(nS)} ({\mathcal A}, \eff)}
      {\lumi \ \cdot \Delta \pt \cdot \Delta y }\,,
\end{align}

upon normalization by the integrated luminosity of the dataset,
$\mathcal{L}$, and by the bin widths, $\Delta \pt$ and $\Delta y$, of the $\PgU$
transverse momentum and rapidity.

The $\PgUa$, $\PgUb$, and $\PgUc$ yields
are extracted via an extended unbinned maximum likelihood fit.
The measured mass-lineshape of each $\PgU$ state
is parameterized by a ``Crystal Ball'' (CB) function~\cite{bib-crystalball};
this is a Gaussian resolution function with the low side tail replaced with a power law describing FSR.
The resolution, given by the Gaussian standard deviation, is a free parameter in the fit but is constrained to scale with the ratios of the resonance masses. The FSR tail is fixed to the MC shape.
Since the three resonances overlap in the measured dimuon mass, we fit the
three $\PgU(nS)$ states simultaneously. Therefore, the probability distribution function (PDF)
describing the signal
consists of three CB functions.  The mass of the $\PgUa$ is a free parameter
in the fit, to accommodate a possible bias in the momentum scale
calibration.  The number of free parameters is reduced by fixing the
$\PgUb$ and $\PgUc$ mass differences, relative to the $\PgUa$, to their world
average values~\cite{pdg}.
A second-order polynomial is chosen to
describe the background in the 8--14\GeVcc mass-fit range.

\begin{figure*}
  \centering
  \includegraphics[angle=0,width=0.9\textwidth]{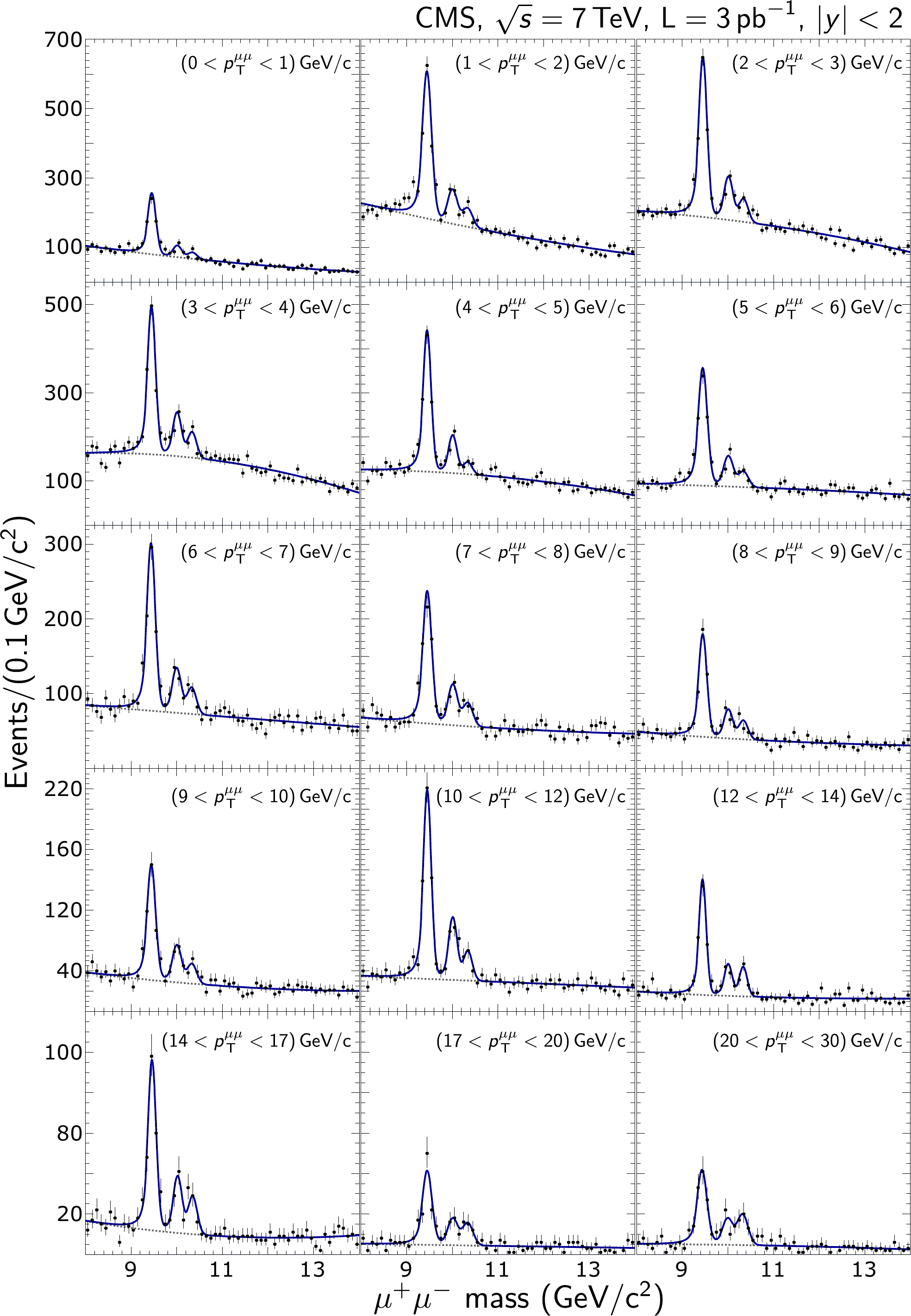}
  \caption{Fit to the dimuon invariant-mass distribution in the specified \pt regions for $|y|<2$, before accounting for acceptance and efficiency. The solid line shows the result of the fit described in the text, with the dashed line representing the background component.}
  \label{fig:massFit-rawPt}
\end{figure*}

The fit to the dimuon invariant-mass spectrum, before
accounting for acceptance and efficiencies, is shown in Fig.~\ref{fig:massFit-raw}
for the $\PgU$ transverse momentum interval $\pt<30\GeVc$,
and for the fifteen \pt intervals used for the $\PgUa$ differential cross-section measurement in Fig.~\ref{fig:massFit-rawPt}.
The observed $\PgU(nS)$ signal yields are reported in Table~\ref{tab:raw_yields}.
The width of the \pt intervals chosen for each resonance reflects the corresponding available signal statistics.
In all cases the quoted uncertainty is statistical.  As shown in  Table~\ref{tab:raw_yields}, for each resonance the sum of the yields in each \pt interval is consistent with the yield determined from a fit to the entire \pt range.
\begin{table}[t]
  \centering
  \caption{The uncorrected $\PgU$ signal yield, fit quality (normalized $\chi^2$, obtained by comparing the fit PDF and the binned data; the number of degrees of freedom is 112), and average weight $\langle w\rangle$ in \pt intervals for $|y|<2$. The mean of the \pt distribution in each interval is also given.\newline}
  {\begin{tabular}{l r@{--}l r@{.}l r@{.}l r@{$\pm$}l r@{.}l }
\hline
&\multicolumn{4}{c}{\pt (\!\GeVc)} & \multicolumn{2}{c}{fit} & \multicolumn{2}{c}{signal} &\multicolumn{2}{c}{} \\
&\multicolumn{2}{c}{range} & \multicolumn{2}{c}{mean}   & \multicolumn{2}{c}{$\chi^2$} &  \multicolumn{2}{c}{yield} & \multicolumn{2}{c}{$\langle w\rangle^{-1}$} \\
\hline
$\PgUa$
&  0 &  1 &  0&7   &   1&1   &  427&34 & 0&44  \\
&  1 &  2 &  1&5   &   1&7   & 1153&54 & 0&41  \\	 
&  2 &  3 &  2&5   &   1&1   & 1154&53 & 0&36  \\	 
&  3 &  4 &  3&5   &   1&3   &  806&46 & 0&30  \\	 
&  4 &  5 &  4&5   &   1&0   &  769&43 & 0&28  \\	 
&  5 &  6 &  5&5   &   1&1   &  716&40 & 0&28  \\	 
&  6 &  7 &  6&5   &   1&2   &  578&37 & 0&28  \\	 
&  7 &  8 &  7&5   &   1&3   &  477&33 & 0&30  \\	 
&  8 &  9 &  8&5   &   1&1   &  344&26 & 0&34  \\	 
&  9 & 10 &  9&5   &   1&1   &  286&24 & 0&37  \\	 
& 10 & 12 & 10&9   &   1&1   &  449&27 & 0&41  \\	 
& 12 & 14 & 12&9   &   1&3   &  246&19 & 0&45  \\	 
& 14 & 17 & 15&4   &   1&2   &  208&18 & 0&50  \\	 
& 17 & 20 & 18&3   &   0&8   &  105&13 & 0&54  \\	 
& 20 & 30 & 23&3   &   0&8   &  109&13 & 0&60  \\	 
& \multicolumn{6}{l}{sum}           & 7825&133 \\     
& \multicolumn{6}{l}{combined fit}  & 7807&133 \\     
\hline
$\PgUb$
&  0& 2   &  1&3 &  1&7 & 368&41 & 0&47 \\
&  2& 4   &  2&9 &  1&3 & 591&50 & 0&40 \\                                                       
&  4& 6   &  4&9 &  0&9 & 416&40 & 0&32 \\                                                       
&  6& 9   &  7&3 &  1&1 & 424&38 & 0&33 \\                                                       
&  9&12   & 10&3 &  1&1 & 257&25 & 0&41 \\                                                       
& 12&16   & 13&6 &  1&3 & 121&16 & 0&46 \\                                                       
& 16&20   & 17&7 &  1&0 &  63&11 & 0&55 \\                                                       
& 20&30   & 22&5 &  0&8 &  39& 9 & 0&60 \\                                                       
& \multicolumn{6}{l}{sum}           &2279&91 &\multicolumn{2}{l}{}\\                                                  
& \multicolumn{6}{l}{combined fit}  &2270&91 &\multicolumn{2}{l}{}\\
\hline
$\PgUc$
&  0& 3   &  1&8 &  1&5  & 397&51 & 0&47 \\                                                       
&  3& 6   &  4&3 &  1&0  & 326&47 & 0&37 \\                                                       
&  6& 9   &  7&3 &  1&1  & 264&36 & 0&35 \\                                                       
&  9&14   & 11&0 &  1&2  & 207&25 & 0&43 \\                                                       
& 14&20   & 16&3 &  1&2  &  83&14 & 0&52 \\                                                       
& 20&30   & 23&4 & 0&8  & 49&10 & 0&61 \\                                  
& \multicolumn{6}{l}{sum}            &  1324&84  &\multicolumn{2}{l}{}\\
& \multicolumn{6}{l}{combined fit}   &  1318&84  &\multicolumn{2}{l}{}\\                           
\hline
\end{tabular}

}
  \label{tab:raw_yields}
\end{table}
Given the significant $\eta$ and \pt dependencies of the efficiencies
and acceptances of the muons from $\PgU(nS)$ decays, we correct
for them on a candidate-by-candidate basis before performing the mass
fit to obtain $N^{\rm{fit}}$ in Eq.~(\ref{eqn:xsection}).
Specifically: an $\PgU$ candidate reconstructed with \pt and $y$ from
muons with $\pt^{\mu_{1,2}}$ and $\eta^{\mu_{1,2}}$ is corrected with
a weight
\begin{equation}
w \equiv w_{\rm acc} \cdot w_{\rm track} \cdot w_{\rm id} \cdot w_{\rm trig} \cdot w_{\rm misc}
\label{eqn:weight}
\end{equation}
where the factors are:
 (i) acceptance, $w_{\rm acc}=1/\acc^{\PgU} (\pt ,y)$;
 (ii) tracking,  $w_{\rm track} = 1/\eff_{\rm track}^2$;
 (iii) identification,  $w_{\rm id}=1/\left[\eff_{\rm id}(\pt^{\mu_1} ,\eta^{\mu_1}) \cdot \eff_{\rm id}(\pt^{\mu_2} ,\eta^{\mu_2})\right]$;
 (iv) trigger, $w_{\rm trig}=1/\left[\eff_{\rm trig}(\pt^{\mu_1} ,\eta^{\mu_1}) \cdot \eff_{\rm trig}(\pt^{\mu_2} ,\eta^{\mu_2})\right]$; and
 (v) additional selection criteria, $w_{\rm misc}$, including the efficiency of the vertex selection criteria.
The acceptance depends on the resonance mass;
the $\PgUc$ gives rise to higher-momenta muons which results in a roughly 10\% larger acceptance for the $\PgUc$ than for the $\PgUa$.
Consequently, the corrected yield for each of the $\PgU(nS)$ resonances is obtained from a fit in which the corresponding $\PgU(nS)$ acceptance is employed.

We determine the $\PgU(nS)$ differential cross section using the above procedure separately for each polarization scenario. The results are summarized in Table~\ref{tab:cross-section-1y}.
We also divide the data into two ranges of rapidity, $|y|<1$ and $1<|y|<2$, and repeat the fits to obtain the $\PgU(nS)$ differential cross sections reported in Table~\ref{tab:cross-section-2y}. The integrated cross section for each resonance is obtained from the corresponding sum of the differential cross sections.
The results for the $\PgUa$ \pt-integrated, rapidity-differential cross section are shown in Table~\ref{tab:cross-section-rapdiff}.

\begin{table}[!htp]
  \centering
  \caption{The product of the $\PgU(nS)$ production cross sections, $\sigma$, and the dimuon branching fraction, ${\cal{B}}$,  measured in \pt bins for $|y|<2$, with the assumption of unpolarized production. The statistical uncertainty (stat.), the sum of the systematic uncertainties in quadrature $(\Sigma_{\rm syst.})$, and the total uncertainty ($\Delta \sigma$; including stat., $\sum_{\rm syst.}$, and luminosity terms) are quoted as relative uncertainties in percent. Values in parentheses denote the negative part of the asymmetric uncertainty. The fractional change in percent of the cross section is shown for four polarization scenarios: fully-longitudinal (L) and fully-transverse (T) in the helicity (HX) and Collins-Soper (CS) frames. \newline}
  \begin{tabular}{c c c c c |  c c c c} \hline
\pt & $\sigma\cdot{\cal{B}}$  & 	stat. &$\sum_{\rm{syst.}}$ & $\Delta\sigma$ & HX-T & HX-L & CS-T & CS-L \\
 (\!\GeVc)  & (nb)  & (\%)  & (\%)  &  (\%)  & (\%)  & (\%)  & (\%) & (\%)  \\
\hline 
\multicolumn{2}{c}{} & \multicolumn{7}{l}{\PgUa \qquad\qquad\quad $|y|<2$} \\ \hline
 0--30	&7.37 & 	${ 1.8}$ &$   8\,(   6)$ & $  14\,(  13)$ &   +16 &   -22 &   +13 &   -16\\ 
 0-- 1	&0.30 & 	${   8}$ &$  10\,(   7)$ & $  17\,(  15)$ &   +16 &   -22 &   +17 &   -23\\ 
 1-- 2	&0.90 & 	${   5}$ &$   9\,(   6)$ & $  15\,(  14)$ &   +16 &   -20 &   +19 &   -24\\ 
 2-- 3	&1.04 & 	${   5}$ &$   8\,(   6)$ & $  14\,(  13)$ &   +15 &   -20 &   +19 &   -24\\ 
 3--4 	&0.88 & 	${   6}$ &$   9\,(   7)$ & $  15\,(  14)$ &   +18 &   -23 &   +18 &   -23\\ 
 4--5 	&0.90 & 	${   6}$ &$   8\,(   6)$ & $  15\,(  14)$ &   +18 &   -23 &   +16 &   -21\\ 
 5--6 	&0.82 & 	${   6}$ &$   8\,(   6)$ & $  15\,(  14)$ &   +17 &   -23 &   +13 &   -19\\ 
 6--7 	&0.64 & 	${   7}$ &$   8\,(   5)$ & $  15\,(  14)$ &   +17 &   -22 &   +11 &   -16\\ 
 7--8 	&0.51 & 	${   7}$ &$   8\,(   6)$ & $  15\,(  14)$ &   +16 &   -22 &    +7 &   -10\\ 
 8--9 	&0.33 & 	${   8}$ &$   8\,(   6)$ & $  16\,(  14)$ &   +16 &   -22 &    +4 &    -5\\ 
 9--10	&0.25 & 	${   8}$ &$   9\,(   6)$ & $  16\,(  15)$ &   +15 &   -21 &    +2 &    -1\\ 
10--12	&0.36 & 	${   6}$ &$   8\,(   5)$ & $  15\,(  14)$ &   +15 &   -21 &    -1 &    +3\\ 
12--14	&0.18 & 	${   8}$ &$   9\,(   5)$ & $  16\,(  14)$ &   +15 &   -20 &    -3 &    +7\\ 
14--17	&0.14 & 	${   9}$ &$  10\,(   6)$ & $  17\,(  15)$ &   +14 &   -19 &    -4 &    +9\\ 
17--20	&0.06 & 	${  12}$ &$  10\,(   6)$ & $  19\,(  17)$ &   +13 &   -18 &    -4 &   +10\\ 
20--30	&0.06 & 	${  12}$ &$  10\,(   6)$ & $  19\,(  17)$ &   +12 &   -17 &    -4 &   +10\\ 
\hline \multicolumn{2}{c}{} & \multicolumn{7}{l}{\PgUb \qquad\qquad\quad $|y|<2$} \\ \hline
 0--30	&1.90 & 	${ 4.2}$ &$   9\,(   6)$ & $  15\,(  13)$ &   +14 &   -19 &   +12 &   -15\\ 
 0--2 	&0.25 & 	${  12}$ &$  11\,(   9)$ & $  20\,(  19)$ &   +14 &   -19 &   +17 &   -22\\ 
 2--4 	&0.48 & 	${   8}$ &$  12\,(  10)$ & $  18\,(  17)$ &   +12 &   -17 &   +18 &   -23\\ 
 4--6 	&0.41 & 	${  10}$ &$  10\,(   8)$ & $  18\,(  17)$ &   +16 &   -22 &   +15 &   -20\\ 
 6--9 	&0.41 & 	${   9}$ &$  10\,(   7)$ & $  17\,(  16)$ &   +15 &   -21 &    +9 &   -13\\ 
 9--12	&0.21 & 	${  10}$ &$   9\,(   6)$ & $  17\,(  16)$ &   +14 &   -20 &    +1 &    -0\\ 
12--16	&0.09 & 	${  13}$ &$  10\,(   7)$ & $  20\,(  19)$ &   +14 &   -19 &    -2 &    +6\\ 
16--20	&0.04 & 	${  18}$ &$  11\,(   8)$ & $  24\,(  23)$ &   +12 &   -18 &    -4 &    +9\\ 
20--30	&0.02 & 	${  23}$ &$  20\,(  18)$ & $  32\,(  32)$ &   +12 &   -17 &    -5 &   +11\\ 
\hline \multicolumn{2}{c}{} & \multicolumn{7}{l}{\PgUc \qquad\qquad\quad $|y|<2$} \\ \hline
 0--30	&1.02 & 	${ 6.7}$ &$  11\,(   8)$ & $  17\,(  15)$ &   +14 &   -19 &   +10 &   -13\\ 
 0--3	&0.26 & 	${  14}$ &$  10\,(   8)$ & $  21\,(  19)$ &   +13 &   -18 &   +16 &   -22\\ 
 3--6	&0.29 & 	${  14}$ &$  18\,(  17)$ & $  26\,(  25)$ &   +13 &   -18 &   +16 &   -21\\ 
 6--9	&0.24 & 	${  14}$ &$  11\,(   8)$ & $  21\,(  19)$ &   +15 &   -20 &   +10 &   -13\\ 
 9--14	&0.16 & 	${  12}$ &$  10\,(   8)$ & $  19\,(  18)$ &   +15 &   -20 &    -1 &    +2\\ 
14--20	&0.05 & 	${  17}$ &$  11\,(   8)$ & $  23\,(  22)$ &   +13 &   -18 &    -4 &    +9\\ 
20--30	&0.03 & 	${  20}$ &$  12\,(   9)$ & $  26\,(  25)$ &   +11 &   -16 &    -4 &    +9\\ 
\hline
\end{tabular}

\label{tab:cross-section-1y}
\end{table}

\begin{table}[!htp]
  \centering
  \caption{The product of the $\PgU(nS)$ production cross sections, $\sigma$, and the dimuon branching fraction, ${\cal{B}}$,  measured in \pt bins for  $|y|<1$ and $1<|y|<2$, with the assumption of unpolarized production. The statistical uncertainty (stat.), the sum of the systematic uncertainties in quadrature $(\Sigma_{\rm syst.})$, and the total uncertainty ($\Delta \sigma$; including stat., $\sum_{\rm syst.}$, and luminosity terms) are quoted as relative uncertainties in percent. Values in parentheses denote the negative part of the asymmetric uncertainty. The fractional change in percent of the cross section is shown for four polarization scenarios: fully-longitudinal (L) and fully-transverse (T) in the helicity (HX) and Collins-Soper (CS) frames. \newline}
  \begin{tabular}{c c c c c |  c c c c} \hline
\pt & $\sigma\cdot{\cal{B}}$  & 	stat. &$\sum_{\rm{syst.}}$ & $\Delta\sigma$ & HX-T & HX-L & CS-T & CS-L \\
 (\!\GeVc)  & (nb)  & (\%)  & (\%)  &  (\%)  & (\%)  & (\%)  & (\%) & (\%)  \\
\hline \multicolumn{2}{c}{} & \multicolumn{7}{l}{\PgUa \qquad\qquad\quad $|y|<1$} \\ \hline
 0--30	&4.03 & 	${ 1.3}$ &$   8\,(   6)$ & $  14\,(  12)$ &   +16 &   -22 &   +13 &   -16\\ 
 0--2	&0.70 & 	${   5}$ &$   9\,(   7)$ & $  15\,(  14)$ &   +14 &   -19 &   +18 &   -24\\ 
 2--5	&1.54 & 	${   4}$ &$  10\,(   9)$ & $  15\,(  15)$ &   +14 &   -20 &   +18 &   -23\\ 
 5--8	&1.02 & 	${   5}$ &$   7\,(   6)$ & $  14\,(  13)$ &   +18 &   -23 &    +8 &   -12\\ 
 8--11	&0.44 & 	${   6}$ &$   7\,(   5)$ & $  15\,(  14)$ &   +18 &   -23 &    -1 &    +2\\ 
11--15	&0.23 & 	${   7}$ &$   8\,(   5)$ & $  15\,(  14)$ &   +18 &   -23 &    -4 &   +10\\ 
15--30	&0.11 & 	${   9}$ &$   8\,(   6)$ & $  16\,(  15)$ &   +15 &   -20 &    -5 &   +12\\ 
\hline \multicolumn{2}{c}{} &  \multicolumn{7}{l}{{\PgUb \qquad\qquad\quad $|y|<1$}} \\ \hline
 0--30	&1.03 & 	${ 2.9}$ &$   9\,(   6)$ & $  15\,(  13)$ &   +14 &   -19 &   +12 &   -15\\ 
 0--3	&0.29 & 	${  10}$ &$  17\,(  16)$ & $  22\,(  21)$ &   +10 &   -14 &   +17 &   -22\\ 
 3--7	&0.41 & 	${  10}$ &$  16\,(  15)$ & $  21\,(  21)$ &   +13 &   -18 &   +14 &   -19\\ 
 7--11	&0.22 & 	${  11}$ &$   9\,(   7)$ & $  18\,(  17)$ &   +17 &   -22 &    +1 &    -2\\ 
11--15	&0.06 & 	${  16}$ &$   9\,(   6)$ & $  21\,(  20)$ &   +17 &   -22 &    -4 &    +8\\ 
15--30	&0.04 & 	${  17}$ &$   9\,(   7)$ & $  22\,(  21)$ &   +14 &   -20 &    -5 &   +11\\ 
\hline \multicolumn{2}{c}{}  &  \multicolumn{7}{l}{\PgUc \qquad\qquad\quad $|y|<1$} \\ \hline
 0--30 &0.59 & 	${ 4.8}$ &$  11\,(   8)$ & $  16\,(  15)$ &   +14 &   -19 &   +10 &   -13\\ 
 0--7  &0.38 & 	${  11}$ &$  25\,(  24)$ & $  30\,(  29)$ &   +11 &   -16 &   +14 &   -19\\ 
 7--12 &0.15 & 	${  15}$ &$  10\,(   8)$ & $  21\,(  20)$ &   +16 &   -22 &    +1 &    -1\\ 
12--30 &0.07 & 	${  14}$ &$  10\,(   8)$ & $  20\,(  20)$ &   +15 &   -21 &    -4 &   +10\\ 
\hline \multicolumn{2}{c}{} & \multicolumn{7}{l}{\PgUa \quad\qquad\quad $1<|y|<2$ } \\ \hline
 0--30 &3.55 & 	${ 1.2}$ &$   8\,(   6)$ & $  14\,(  12)$ &   +16 &   -22 &   +13 &   -16\\ 
 0--2  &0.55 & 	${   7}$ &$  11\,(   9)$ & $  17\,(  16)$ &   +18 &   -24 &   +18 &   -23\\ 
 2--5  &1.39 & 	${   4}$ &$   9\,(   7)$ & $  15\,(  14)$ &   +20 &   -25 &   +18 &   -23\\ 
 5--8  &0.97 & 	${   5}$ &$   9\,(   5)$ & $  15\,(  13)$ &   +16 &   -22 &   +14 &   -18\\ 
 8--11 &0.37 & 	${   7}$ &$  10\,(   6)$ & $  16\,(  14)$ &   +13 &   -19 &    +6 &    -8\\ 
11--15 &0.18 & 	${   8}$ &$  10\,(   6)$ & $  17\,(  15)$ &   +11 &   -17 &     0 &    +1\\ 
15--30 &0.10 & 	${   9}$ &$  11\,(   6)$ & $  18\,(  16)$ &   +10 &   -16 &    -3 &    +6\\ 
\hline \multicolumn{2}{c}{} &  \multicolumn{7}{l}{\PgUb \quad\qquad\quad $1<|y|<2$ } \\ \hline
 0--30 &0.93 & 	${ 3.0}$ &$   9\,(   6)$ & $  15\,(  13)$ &   +14 &   -19 &   +12 &   -15\\ 
 0--3  &0.21 & 	${  15}$ &$  24\,(  23)$ & $  30\,(  29)$ &   +17 &   -23 &   +17 &   -23\\ 
 3--7  &0.44 & 	${   9}$ &$  12\,(   8)$ & $  18\,(  17)$ &   +17 &   -22 &   +17 &   -22\\ 
 7--11 &0.19 & 	${  12}$ &$  11\,(   8)$ & $  20\,(  18)$ &   +13 &   -18 &    +9 &   -12\\ 
11--15 &0.06 & 	${  17}$ &$  11\,(   7)$ & $  23\,(  21)$ &   +11 &   -17 &    +1 &     0\\ 
15--30 &0.03 & 	${  21}$ &$  13\,(   9)$ & $  27\,(  26)$ &   +10 &   -16 &    -3 &    +7\\ 
\hline \multicolumn{2}{c}{} & \multicolumn{7}{l}{\PgUc \quad\qquad\quad $1<|y|<2$ } \\ \hline
 0--30	&0.40 & 	${ 4.9}$ &$  11\,(   8)$ & $  16\,(  15)$ &   +14 &   -19 &   +10 &   -13\\ 
 0--7	&0.24 & 	${  18}$ &$  29\,(  27)$ & $  36\,(  35)$ &   +16 &   -22 &   +17 &   -22\\ 
 7--12	&0.10 & 	${  22}$ &$  13\,(  10)$ & $  28\,(  27)$ &   +13 &   -18 &   +10 &   -13\\ 
12--30	&0.06 & 	${  17}$ &$  11\,(   8)$ & $  23\,(  22)$ &   +10 &   -15 &    -2 &    +5\\ 
\hline
\end{tabular}

  \label{tab:cross-section-2y}
\end{table}

\begin{table}[h!]
  \centering
  \caption{The product of the $\PgUa$ production cross section, $\sigma$, and the dimuon branching fraction, ${\cal{B}}$,  measured in rapidity bins and integrated over the \pt range  $\pt^\PgU<30\GeVc$, with the assumption of unpolarized production. The statistical uncertainty (stat.), the sum of the systematic uncertainties in quadrature $(\Sigma_{\rm syst.})$, and the total uncertainty ($\Delta \sigma$; including stat., $\sum_{\rm syst.}$, and luminosity terms) are quoted as relative uncertainties in percent. Values in parentheses denote the negative part of the asymmetric uncertainty. The fractional change in percent of the cross section is shown for four polarization scenarios: fully-longitudinal (L) and fully-transverse (T) in the helicity (HX) and Collins-Soper (CS) frames. \newline}
  \renewcommand{\arraystretch}{1.2}
  \begin{tabular}{c c c c c |  c c c c} \hline
$|y|$ & $\sigma\cdot{\cal{B}}$  & 	stat. &$\sum_{\rm{syst.}}$ & $\Delta\sigma$ & HX-T & HX-L & CS-T & CS-L \\
  & (nb)  & (\%)  & (\%)  &  (\%)  & (\%)  & (\%)  & (\%) & (\%)  \\
\hline \multicolumn{2}{c}{} & \multicolumn{7}{l}{\PgUa \quad\qquad\quad $\pt<30\GeVc$} \\ \hline
0.0--2.0 &7.61 & 	${ 1.8}$ &$   8\,(   6)$ & $  14\,(  13)$ &   +16 &   -22 &   +13 &   -16\\ 
0.0--0.4 &1.62 & 	${   3}$ &$   8\,(   6)$ & $  14\,(  13)$ &   +15 &   -19 &   +13 &   -17\\ 
0.4--0.8 &1.52 & 	${   4}$ &$   9\,(   8)$ & $  15\,(  14)$ &   +17 &   -22 &   +11 &   -15\\ 
0.8--1.2 &1.77 & 	${   4}$ &$   9\,(   7)$ & $  14\,(  13)$ &   +16 &   -22 &    +9 &   -12\\ 
1.2--1.6 &1.47 & 	${   4}$ &$   9\,(   7)$ & $  15\,(  13)$ &   +17 &   -23 &   +12 &   -16\\ 
1.6--2.0 &1.23 & 	${   4}$ &$  11\,(   7)$ & $  16\,(  14)$ &   +18 &   -23 &   +20 &   -24\\ 
\hline
\end{tabular}

  \label{tab:cross-section-rapdiff}
\end{table}

\section{Systematic uncertainties}
\label{sec:systematics}

Systematic uncertainties are described in this section, together with the methods used
in their determination. We give a representative value for each uncertainty in parentheses.

We determine the cross section using acceptance maps corresponding to five different polarization scenarios, expected to represent extreme cases.
The values of the cross section obtained vary by about 20\%.
The variations depend on \pt thus affecting the shapes of the \pt spectrum.

The statistical uncertainties on the acceptance and efficiencies
-- single-muon trigger and muon ID, quality criteria, tracking and vertex quality --
give rise to systematic uncertainties in the cross-section measurement.
We vary the dimuon event weights in the fit coherently by $\pm 1 \sigma {\rm (stat.)}$.
The muon identification and trigger efficiencies are varied coherently when estimating the associated systematic uncertainties (8\%).

The selection criteria requiring the muons to be consistent with emanating from the same primary vertex are fully efficient.  This has been confirmed in data and simulation.
The selection of one candidate per event using the largest vertex probability also has an efficiency consistent with unity.
We assign an uncertainty (0.2\%) from the frequency of occurrence of signal candidates in the data that are rejected by the largest vertex probability requirement but pass all the remaining selection criteria.
The muon charge misassignment is estimated to be less than 0.01\%~\cite{muoncharge} and contributes a negligible uncertainty.

Final-state radiation is incorporated into the simulation using the \PHOTOS algorithm.
To estimate the systematic uncertainty associated with this procedure, the acceptance is calculated without FSR
and 20\% of the difference is taken as the uncertainty based on a study in Ref.~\cite{bib-photos2} (0.8\%).

The definition of acceptance used in this analysis requires that the muons from the $\PgU$ decay produce reconstructible tracks.
The kinematic selection is applied to the reconstructed \pt and $\eta$ values
of these tracks.  Uncertainties on the measurement of track parameters
also affect the acceptance as a systematic uncertainty.  The dominant
uncertainty is associated with the measurement of the track transverse momentum.
The acceptance is sensitive to biases in track momentum and to
differences in resolution between the simulated and measured
distributions.  The magnitude of these effects is quantified by
comparing measurements of resonance mass and width between
simulation and data~\cite{bib-trackermomentum}.  To determine the
effect on the $\PgU$ acceptance, we introduce a track \pt bias of
0.2\%, chosen to be four times
the maximum momentum scale residual bias after calibration (0.3\%).
We also vary the transverse momentum resolution by $\pm 10\%$,
corresponding to the uncertainty in the resolution
measurement using $\JPsi$, and recalculate the acceptance
map (0.1\%).

Imperfect knowledge of the production \pt spectrum of the $\PgU$ resonances at $\sqrts=7\TeV$ contributes a systematic uncertainty. The $\PgU$ MC sample used for the acceptance calculation, Eq.~(\ref{eqn:acc}), was generated flat in \pt, whereas the \pt spectrum in the data peaks at a few \GeVc, and behaves as a power law above 5\GeVc. To study the effect of this difference, we have re-weighted the sample in \pt to more closely describe the expected distribution in data based on a fit to the spectrum obtained from \PYTHIA (1\%).

The distribution of the $z$ position of the \Pp\Pp interaction point influences the acceptance. We have produced MC
samples of $\PgU(nS)$ at different positions along the beam line, between $-10$ and $+10$\cm with respect to the center of the nominal collision region (1\%).

High-statistics MC simulations were performed to compare T\&P single-muon and dimuon efficiencies to the actual MC 
values for both the $\PgU$ and $\JPsi$, see Figs.~\ref{fig:TurnOnMuId} and~\ref{fig:TurnOnTrig}. 
The differences and their associated uncertainties are taken as a source of systematic uncertainty. 
The contributions are:  possible bias in the T\&P technique (0.1\%),
differences in the $\JPsi$ and $\PgU$ kinematics (1\%), and taking the product of single-muon
efficiencies as an estimate of the double-muon $\PgU$ efficiencies (1.6\%).

MC trials of the fitter demonstrate that it is consistent with providing an
unbiased estimate of the yield of each resonance, its mass, and the mass resolution (1\%).
A systematic variation may arise from differences between the dimuon invariant-mass distribution
in the data and in the PDFs chosen for the signal and background
components in the fit.
We consider the following variations in the signal PDF. As the CB parameters which describe the radiative tail of each resonance are fixed from
MC simulation in the nominal fit to the data,
we vary the CB parameters by three times their uncertainties (3\%).
We also remove the resonance mass difference constraint in the \pt integrated fit (0.6\%).
We vary the background PDF by replacing the polynomial by a linear function, while restricting the fit to the mass range 8--12\GeVcc (3\% when fitting the full \pt and $y$ ranges, varying with differential interval).

The determination of the integrated luminosity normalization is made with an uncertainty of 11\%~\cite{bib-lumi-pas}.
The relative systematic uncertainties from each source are summarized in Table~\ref{tab:syst-xsec-1y} for the full rapidity range, for two rapidity ranges in Table~\ref{tab:syst-xsec-2y},
and for five rapidity ranges in Table~\ref{tab:syst-xsec-rapdiff}.
The largest sources of systematic uncertainty arise from the statistical precision of the T\&P determination of the efficiencies from the data and from the luminosity normalization which dominates.

\begin{table*}[htp]
  \centering
  \caption{Relative values of the systematic uncertainties on the $\PgU(nS)$ production cross sections times the dimuon branching fraction, in  \pt intervals for $|y|<2$, assuming unpolarized production, in percent. The following abbreviations are used: $A$, $\eff_{\rm trig,id}$, $S_p$, $A_{\pt}$, $A_{\rm vtx}$, $A_{\rm FSR}$, ${\rm T\&P}$, and $\eff_{\JPsi,\PgU}$, for the systematic uncertainties arising from imperfect knowledge of the acceptance, trigger and muon identification efficiencies, momentum scale, the production \pt spectrum, the efficiency of the vertex quality criterion, the modeling of FSR, the T\&P method, and the bias from using the $\JPsi$ to determine single-muon efficiencies rather than  the $\PgU$. The uncertainties associated with the background PDF are in the column labeled {\sc bg}, while the signal PDF, the fitter, tracking efficiency, and effects arising from the efficiency binning are combined in the column labeled {add}. Values in parentheses denote the negative part of the asymmetric uncertainty. The luminosity uncertainty of 11\% is not included in the table.\newline}
  \begin{tabular}{c|c|c|c|c|c|c|c|c|c|c} \hline
$\pt$ (\!\GeVc)  &         $A$ & $\eff_{\rm{trig,id}}$ &     $S_{p}$ & $A_{\pt}$ & $A_{\rm vtx}$ & $A_{\rm FSR}$ &        T\&P & $\eff_{\JPsi,\PgU}$ & {\sc bg} &        add.\\
\hline \multicolumn{1}{c}{} & \multicolumn{10}{l}{\PgUa \qquad $|y|<2$ \qquad uncertainties are in percent } \\ \hline
 0--30	&$ 0.5\,( 0.5)$ &$ 7.5\,( 4.6)$ &$ 0.3\,( 0.3)$ & $ 0.6$ & $ 0.7$ & $ 0.7$ & $ 0.0$ & $ 0.9$ & $ 0.5$ &$ 3.0$ \\ 
 0--1	&$ 0.4\,( 0.4)$ &$ 8.3\,( 5.4)$ &$ 0.1\,( 0.1)$ & $ 0.2$ & $ 1.1$ & $ 0.8$ & $ 0.5$ & $ 0.8$ & $ 3.4$ &$ 3.1$ \\ 
 1--2	&$ 0.4\,( 0.4)$ &$ 7.8\,( 5.2)$ &$ 0.2\,( 0.2)$ & $ 0.6$ & $ 0.6$ & $ 0.7$ & $ 0.2$ & $ 1.1$ & $ 1.8$ &$ 3.0$ \\ 
 2--3	&$ 0.5\,( 0.5)$ &$ 7.3\,( 4.7)$ &$ 0.6\,( 0.6)$ & $ 0.3$ & $ 0.3$ & $ 0.8$ & $ 0.1$ & $ 1.1$ & $ 1.5$ &$ 3.0$ \\ 
 3--4	&$ 0.6\,( 0.6)$ &$ 7.3\,( 4.8)$ &$ 0.6\,( 0.6)$ & $ 0.1$ & $ 0.4$ & $ 0.8$ & $ 0.0$ & $ 1.1$ & $ 3.7$ &$ 3.0$ \\ 
 4--5	&$ 0.6\,( 0.6)$ &$ 7.4\,( 4.5)$ &$ 0.4\,( 0.3)$ & $ 0.3$ & $ 0.7$ & $ 0.7$ & $ 0.0$ & $ 0.9$ & $ 2.3$ &$ 3.0$ \\ 
 5--6	&$ 0.6\,( 0.6)$ &$ 7.4\,( 4.3)$ &$ 0.2\,( 0.3)$ & $ 0.5$ & $ 1.0$ & $ 0.7$ & $ 0.0$ & $ 0.7$ & $ 0.5$ &$ 3.0$ \\ 
 6--7	&$ 0.6\,( 0.6)$ &$ 7.4\,( 4.1)$ &$ 0.2\,( 0.3)$ & $ 0.7$ & $ 1.1$ & $ 0.6$ & $ 0.1$ & $ 0.7$ & $ 0.4$ &$ 3.0$ \\ 
 7--8	&$ 0.6\,( 0.6)$ &$ 7.7\,( 4.7)$ &$ 0.1\,( 0.1)$ & $ 1.0$ & $ 0.7$ & $ 0.6$ & $ 0.2$ & $ 0.8$ & $ 1.0$ &$ 3.1$ \\ 
 8--9	&$ 0.6\,( 0.6)$ &$ 7.4\,( 4.2)$ &$ 0.0\,( 0.1)$ & $ 1.2$ & $ 0.7$ & $ 0.5$ & $ 0.0$ & $ 0.7$ & $ 1.0$ &$ 3.0$ \\ 
 9--10	&$ 0.5\,( 0.5)$ &$ 7.8\,( 4.3)$ &$ 0.1\,( 0.0)$ & $ 1.3$ & $ 0.9$ & $ 0.5$ & $ 0.2$ & $ 0.6$ & $ 1.9$ &$ 3.1$ \\ 
10--12	&$ 0.5\,( 0.5)$ &$ 7.4\,( 3.7)$ &$ 0.1\,( 0.1)$ & $ 1.4$ & $ 0.8$ & $ 0.5$ & $ 0.2$ & $ 0.6$ & $ 0.2$ &$ 3.0$ \\ 
12--14	&$ 0.5\,( 0.4)$ &$ 7.9\,( 4.0)$ &$ 0.2\,( 0.1)$ & $ 1.6$ & $ 0.9$ & $ 0.5$ & $ 0.1$ & $ 0.6$ & $ 0.3$ &$ 3.1$ \\ 
14--17	&$ 0.4\,( 0.4)$ &$ 8.5\,( 4.2)$ &$ 0.1\,( 0.1)$ & $ 1.6$ & $ 0.9$ & $ 0.5$ & $ 0.3$ & $ 0.6$ & $ 2.2$ &$ 3.1$ \\ 
17--20	&$ 0.4\,( 0.4)$ &$ 8.9\,( 4.4)$ &$ 0.1\,( 0.1)$ & $ 1.8$ & $ 0.8$ & $ 0.4$ & $ 0.5$ & $ 0.7$ & $ 0.1$ &$ 3.6$ \\ 
20--30	&$ 0.3\,( 0.3)$ &$ 8.9\,( 4.3)$ &$ 0.1\,( 0.1)$ & $ 1.6$ & $ 0.7$ & $ 0.5$ & $ 0.3$ & $ 0.6$ & $ 0.1$ &$ 3.5$ \\ 
\hline \multicolumn{1}{c}{} & \multicolumn{10}{l}{\PgUb \qquad $|y|<2$} \\ \hline
 0--30	&$ 0.6\,( 0.6)$ &$ 8.3\,( 4.9)$ &$ 0.3\,( 0.3)$ & $ 0.7$ & $ 0.8$ & $ 0.8$ & $ 0.0$ & $ 1.0$ & $ 1.9$ &$ 3.2$ \\ 
 0--2	&$ 0.5\,( 0.5)$ &$ 8.3\,( 5.2)$ &$ 0.2\,( 0.2)$ & $ 0.5$ & $ 0.6$ & $ 0.8$ & $ 0.4$ & $ 0.6$ & $ 6.8$ &$ 3.3$ \\ 
 2--4	&$ 0.7\,( 0.7)$ &$ 8.3\,( 5.4)$ &$ 0.7\,( 0.8)$ & $ 0.2$ & $ 0.3$ & $ 1.0$ & $ 0.1$ & $ 1.5$ & $ 8.0$ &$ 3.3$ \\ 
 4--6	&$ 0.8\,( 0.7)$ &$ 7.9\,( 4.7)$ &$ 0.4\,( 0.4)$ & $ 0.4$ & $ 1.1$ & $ 0.8$ & $ 0.0$ & $ 0.9$ & $ 5.2$ &$ 3.3$ \\ 
 6--9	&$ 0.7\,( 0.7)$ &$ 8.6\,( 4.8)$ &$ 0.1\,( 0.1)$ & $ 1.0$ & $ 1.2$ & $ 0.7$ & $ 0.2$ & $ 0.9$ & $ 1.7$ &$ 3.5$ \\ 
 9--12	&$ 0.5\,( 0.5)$ &$ 8.4\,( 4.2)$ &$ 0.1\,( 0.1)$ & $ 1.5$ & $ 1.0$ & $ 0.5$ & $ 0.2$ & $ 0.8$ & $ 0.9$ &$ 3.6$ \\ 
12--16	&$ 0.4\,( 0.5)$ &$ 8.8\,( 4.6)$ &$ 0.1\,( 0.1)$ & $ 1.6$ & $ 0.9$ & $ 0.5$ & $ 0.3$ & $ 0.8$ & $ 2.0$ &$ 4.0$ \\ 
16--20	&$ 0.3\,( 0.4)$ &$ 8.3\,( 4.1)$ &$ 0.2\,( 0.1)$ & $ 1.7$ & $ 1.0$ & $ 0.5$ & $ 0.4$ & $ 0.5$ & $ 0.0$ &$ 6.5$ \\ 
20--30	&$ 0.3\,( 0.3)$ &$ 9.1\,( 4.4)$ &$ 0.1\,( 0.1)$ & $ 1.7$ & $ 0.8$ & $ 0.5$ & $ 0.2$ & $ 0.3$ & $ 0.0$ &$17.3$ \\ 
\hline \multicolumn{1}{c}{} & \multicolumn{10}{l}{\PgUc \qquad $|y|<2$} \\ \hline
 0--30	&$ 0.7\,( 0.6)$ &$ 8.6\,( 4.7)$ &$ 0.3\,( 0.3)$ & $ 0.8$ & $ 0.8$ & $ 0.8$ & $ 0.1$ & $ 1.0$ & $ 3.4$ &$ 5.4$ \\ 
 0--3	&$ 0.5\,( 0.5)$ &$ 8.5\,( 4.4)$ &$ 0.4\,( 0.5)$ & $ 0.5$ & $ 0.4$ & $ 0.9$ & $ 0.2$ & $ 0.6$ & $ 1.7$ &$ 5.7$ \\ 
 3--6	&$ 0.9\,( 0.8)$ &$ 9.1\,( 5.4)$ &$ 0.7\,( 0.7)$ & $ 0.3$ & $ 0.9$ & $ 1.0$ & $ 0.0$ & $ 1.7$ & $14.1$ &$ 7.3$ \\ 
 6--9	&$ 0.7\,( 0.7)$ &$ 8.9\,( 4.8)$ &$ 0.2\,( 0.2)$ & $ 1.1$ & $ 1.0$ & $ 0.7$ & $ 0.0$ & $ 1.0$ & $ 2.2$ &$ 5.6$ \\ 
 9--14	&$ 0.5\,( 0.5)$ &$ 7.5\,( 4.1)$ &$ 0.1\,( 0.1)$ & $ 1.5$ & $ 0.8$ & $ 0.5$ & $ 0.3$ & $ 0.7$ & $ 0.4$ &$ 6.1$ \\ 
14--20	&$ 0.4\,( 0.4)$ &$ 8.8\,( 4.5)$ &$ 0.2\,( 0.1)$ & $ 1.7$ & $ 0.8$ & $ 0.5$ & $ 0.3$ & $ 0.6$ & $ 3.4$ &$ 5.9$ \\ 
20--30	&$ 0.3\,( 0.3)$ &$ 8.8\,( 4.1)$ &$ 0.1\,( 0.1)$ & $ 1.6$ & $ 0.8$ & $ 0.5$ & $ 0.5$ & $ 0.5$ & $ 0.3$ &$ 8.3$ \\ 
\hline
\end{tabular}

  \label{tab:syst-xsec-1y}
\end{table*}

\begin{table*}[htp]
  \centering
  \caption{Relative values of the systematic uncertainties on the $\PgU(nS)$ production cross sections times the dimuon branching fraction, in  \pt intervals for $|y|<1$ and $1<|y|<2$, assuming unpolarized production, in percent. The following abbreviations are used: $A$, $\eff_{\rm trig,id}$, $S_p$, $A_{\pt}$, $A_{\rm vtx}$, $A_{\rm FSR}$, ${\rm T\&P}$, and $\eff_{\JPsi,\PgU}$, for the systematic uncertainties arising from imperfect knowledge of the acceptance, trigger and muon identification efficiencies, momentum scale, the production \pt spectrum, the efficiency of the vertex quality criterion, the modeling of FSR, the T\&P method, and the bias from using the $\JPsi$ to determine single-muon efficiencies rather than  the $\PgU$. The uncertainties associated with the background PDF are in the column labeled {\sc bg}, while the signal PDF, the fitter, tracking efficiency, and effects arising from the efficiency binning are combined in the column labeled {add}. Values in parentheses denote the negative part of the asymmetric uncertainty. The luminosity uncertainty of 11\% is not included in the table.\newline}
  \begin{tabular}{c|c|c|c|c|c|c|c|c|c|c} \hline
$\pt$ (\!\GeVc)  &         $A$ & $\eff_{\rm{trig,id}}$ &     $S_{p}$ & $A_{\pt}$ & $A_{\rm vtx}$ & $A_{\rm FSR}$ &        T\&P & $\eff_{\JPsi,\PgU}$ & {\sc bg} &        add.\\
\hline \multicolumn{1}{c}{} & \multicolumn{10}{l}{\PgUa \qquad $|y|<1$ \qquad uncertainties are in percent } \\ \hline
 0-30 &$ 0.5\,( 0.5)$ &$ 7.5\,( 4.6)$ &$ 0.3\,( 0.3)$ & $ 0.6$ & $ 0.7$ & $ 0.7$ & $ 0.0$ & $ 0.9$ & $ 0.5$ &$ 3.0$ \\ 
 0-2  &$ 0.4\,( 0.4)$ &$ 7.7\,( 5.6)$ &$ 0.3\,( 0.3)$ & $ 0.6$ & $ 0.5$ & $ 0.7$ & $ 0.7$ & $ 1.3$ & $ 1.6$ &$ 3.0$ \\ 
 2-5  &$ 0.6\,( 0.6)$ &$ 7.1\,( 5.2)$ &$ 0.7\,( 0.7)$ & $ 0.2$ & $ 0.1$ & $ 0.9$ & $ 0.3$ & $ 1.5$ & $ 6.2$ &$ 3.0$ \\ 
 5-8  &$ 0.7\,( 0.7)$ &$ 6.5\,( 4.4)$ &$ 0.3\,( 0.3)$ & $ 0.8$ & $ 0.8$ & $ 0.7$ & $ 0.4$ & $ 1.0$ & $ 0.3$ &$ 3.0$ \\ 
 8-11 &$ 0.5\,( 0.5)$ &$ 6.4\,( 3.9)$ &$ 0.0\,( 0.0)$ & $ 1.3$ & $ 0.5$ & $ 0.5$ & $ 0.3$ & $ 0.7$ & $ 0.6$ &$ 3.0$ \\ 
11-15 &$ 0.5\,( 0.4)$ &$ 6.6\,( 3.8)$ &$ 0.1\,( 0.1)$ & $ 1.5$ & $ 0.7$ & $ 0.5$ & $ 0.3$ & $ 0.6$ & $ 0.4$ &$ 3.0$ \\ 
15-30 &$ 0.3\,( 0.4)$ &$ 7.1\,( 4.2)$ &$ 0.1\,( 0.2)$ & $ 1.6$ & $ 0.6$ & $ 0.5$ & $ 0.3$ & $ 0.5$ & $ 0.9$ &$ 3.0$ \\ 
\hline \multicolumn{1}{c}{} & \multicolumn{10}{l}{\PgUb \qquad $|y|<1$} \\ \hline
 0-30 &$ 0.6\,( 0.6)$ &$ 8.3\,( 4.9)$ &$ 0.3\,( 0.3)$ & $ 0.7$ & $ 0.8$ & $ 0.8$ & $ 0.0$ & $ 1.0$ & $ 1.9$ &$ 3.2$ \\ 
 0-3  &$ 0.6\,( 0.5)$ &$ 8.2\,( 6.0)$ &$ 0.6\,( 0.6)$ & $ 0.5$ & $ 0.1$ & $ 1.0$ & $ 0.7$ & $ 1.4$ & $13.9$ &$ 3.4$ \\ 
 3-7  &$ 0.8\,( 0.8)$ &$ 7.7\,( 5.2)$ &$ 0.6\,( 0.7)$ & $ 0.4$ & $ 0.6$ & $ 1.0$ & $ 0.4$ & $ 1.5$ & $13.1$ &$ 3.4$ \\ 
 7-11 &$ 0.6\,( 0.6)$ &$ 7.7\,( 4.9)$ &$ 0.1\,( 0.0)$ & $ 1.3$ & $ 0.8$ & $ 0.6$ & $ 0.3$ & $ 1.0$ & $ 1.7$ &$ 3.4$ \\ 
11-15 &$ 0.5\,( 0.5)$ &$ 7.3\,( 4.4)$ &$ 0.1\,( 0.1)$ & $ 1.6$ & $ 0.8$ & $ 0.5$ & $ 0.3$ & $ 0.7$ & $ 1.6$ &$ 3.6$ \\ 
15-30 &$ 0.3\,( 0.4)$ &$ 7.4\,( 4.3)$ &$ 0.1\,( 0.2)$ & $ 1.7$ & $ 0.5$ & $ 0.5$ & $ 0.2$ & $ 0.4$ & $ 1.9$ &$ 4.2$ \\ 
\hline \multicolumn{1}{c}{} & \multicolumn{10}{l}{\PgUc \qquad $|y|<1$} \\ \hline
 0-30 &$ 0.7\,( 0.6)$ &$ 8.6\,( 4.7)$ &$ 0.3\,( 0.3)$ & $ 0.8$ & $ 0.8$ & $ 0.8$ & $ 0.1$ & $ 1.0$ & $ 3.4$ &$ 5.4$ \\ 
 0-7  &$ 0.8\,( 0.8)$ &$ 8.8\,( 5.9)$ &$ 0.6\,( 0.7)$ & $ 0.5$ & $ 0.5$ & $ 1.0$ & $ 0.5$ & $ 1.7$ & $22.9$ &$ 5.4$ \\ 
 7-12 &$ 0.6\,( 0.6)$ &$ 7.6\,( 5.0)$ &$ 0.0\,( 0.0)$ & $ 1.4$ & $ 0.6$ & $ 0.6$ & $ 0.1$ & $ 0.9$ & $ 2.2$ &$ 5.7$ \\ 
12-30 &$ 0.4\,( 0.4)$ &$ 7.1\,( 4.0)$ &$ 0.2\,( 0.1)$ & $ 1.6$ & $ 0.7$ & $ 0.5$ & $ 0.3$ & $ 0.5$ & $ 0.2$ &$ 6.0$ \\ 
\hline \multicolumn{1}{c}{} & \multicolumn{10}{l}{\PgUa \qquad $1<|y|<2$} \\ \hline
 0-30 &$ 0.5\,( 0.5)$ &$ 7.5\,( 4.6)$ &$ 0.3\,( 0.3)$ & $ 0.6$ & $ 0.7$ & $ 0.7$ & $ 0.0$ & $ 0.9$ & $ 0.5$ &$ 3.0$ \\ 
 0-2  &$ 0.4\,( 0.4)$ &$ 8.2\,( 4.6)$ &$ 0.0\,( 0.1)$ & $ 0.3$ & $ 1.1$ & $ 0.6$ & $ 0.2$ & $ 0.6$ & $ 7.3$ &$ 3.0$ \\ 
 2-5  &$ 0.5\,( 0.5)$ &$ 7.7\,( 4.0)$ &$ 0.3\,( 0.3)$ & $ 0.2$ & $ 0.9$ & $ 0.7$ & $ 0.3$ & $ 0.5$ & $ 4.3$ &$ 3.0$ \\ 
 5-8  &$ 0.6\,( 0.6)$ &$ 8.4\,( 4.2)$ &$ 0.1\,( 0.2)$ & $ 0.7$ & $ 1.2$ & $ 0.6$ & $ 0.5$ & $ 0.6$ & $ 0.4$ &$ 3.0$ \\ 
 8-11 &$ 0.6\,( 0.5)$ &$ 8.9\,( 4.4)$ &$ 0.0\,( 0.1)$ & $ 1.3$ & $ 1.1$ & $ 0.5$ & $ 0.6$ & $ 0.7$ & $ 1.7$ &$ 3.1$ \\ 
11-15 &$ 0.4\,( 0.5)$ &$ 9.1\,( 4.2)$ &$ 0.1\,( 0.2)$ & $ 1.6$ & $ 0.9$ & $ 0.5$ & $ 0.8$ & $ 0.6$ & $ 0.1$ &$ 3.0$ \\ 
15-30 &$ 0.4\,( 0.5)$ &$10.6\,( 4.3)$ &$ 0.2\,( 0.3)$ & $ 1.7$ & $ 1.1$ & $ 0.5$ & $ 0.9$ & $ 0.8$ & $ 2.0$ &$ 3.1$ \\ 
\hline \multicolumn{1}{c}{} & \multicolumn{10}{l}{\PgUb \qquad $1<|y|<2$} \\ \hline
 0-30 &$ 0.6\,( 0.6)$ &$ 8.3\,( 4.9)$ &$ 0.3\,( 0.3)$ & $ 0.7$ & $ 0.8$ & $ 0.8$ & $ 0.0$ & $ 1.0$ & $ 1.9$ &$ 3.3$ \\ 
 0-3  &$ 0.5\,( 0.5)$ &$ 7.8\,( 3.8)$ &$ 0.2\,( 0.2)$ & $ 0.3$ & $ 1.2$ & $ 0.7$ & $ 0.2$ & $ 0.5$ & $21.9$ &$ 3.8$ \\ 
 3-7  &$ 0.7\,( 0.7)$ &$ 9.4\,( 4.9)$ &$ 0.3\,( 0.3)$ & $ 0.4$ & $ 1.3$ & $ 0.8$ & $ 0.4$ & $ 0.6$ & $ 5.4$ &$ 3.4$ \\ 
 7-11 &$ 0.6\,( 0.6)$ &$ 9.6\,( 4.8)$ &$ 0.0\,( 0.0)$ & $ 1.2$ & $ 1.1$ & $ 0.6$ & $ 0.7$ & $ 0.8$ & $ 4.4$ &$ 3.5$ \\ 
11-15 &$ 0.5\,( 0.5)$ &$ 9.7\,( 4.8)$ &$ 0.1\,( 0.2)$ & $ 1.6$ & $ 0.9$ & $ 0.5$ & $ 1.1$ & $ 0.7$ & $ 1.5$ &$ 4.6$ \\ 
15-30 &$ 0.4\,( 0.4)$ &$ 9.5\,( 3.8)$ &$ 0.2\,( 0.2)$ & $ 1.7$ & $ 1.3$ & $ 0.4$ & $ 1.0$ & $ 0.7$ & $ 3.4$ &$ 7.2$ \\ 
\hline \multicolumn{1}{c}{} & \multicolumn{10}{l}{\PgUc \qquad $1<|y|<2$} \\ \hline
 0-30 &$ 0.7\,( 0.6)$ &$ 8.6\,( 4.7)$ &$ 0.3\,( 0.3)$ & $ 0.8$ & $ 0.8$ & $ 0.8$ & $ 0.1$ & $ 1.0$ & $ 3.4$ &$ 5.5$ \\ 
 0-7  &$ 0.7\,( 0.6)$ &$ 8.6\,( 2.9)$ &$ 0.4\,( 0.4)$ & $ 0.4$ & $ 1.1$ & $ 0.8$ & $ 0.5$ & $ 0.3$ & $26.5$ &$ 6.5$ \\ 
 7-12 &$ 0.7\,( 0.6)$ &$ 9.3\,( 4.1)$ &$ 0.0\,( 0.0)$ & $ 1.3$ & $ 1.0$ & $ 0.6$ & $ 1.1$ & $ 0.6$ & $ 6.2$ &$ 6.7$ \\ 
12-30 &$ 0.4\,( 0.4)$ &$ 9.4\,( 4.3)$ &$ 0.1\,( 0.2)$ & $ 1.7$ & $ 1.1$ & $ 0.5$ & $ 0.9$ & $ 0.5$ & $ 1.5$ &$ 5.9$ \\ 
\hline
\end{tabular}

\label{tab:syst-xsec-2y}
\end{table*}

\begin{table*}[htp!]
  \centering
  \caption{Relative values of the systematic uncertainties on the $\PgUa$ production cross section times the dimuon branching fraction, in rapidity intervals for $\pt<30\GeVc$, assuming unpolarized production, in percent. The following abbreviations are used: $A$, $\eff_{\rm trig,id}$, $S_p$, $A_{\pt}$, $A_{\rm vtx}$, $A_{\rm FSR}$, ${\rm T\&P}$, and $\eff_{\JPsi,\PgU}$, for the systematic uncertainties arising from imperfect knowledge of the acceptance, trigger and muon identification efficiencies, momentum scale, the production \pt spectrum, the efficiency of the vertex quality criterion, the modeling of FSR, the T\&P method, and the bias from using the $\JPsi$ to determine single-muon efficiencies rather than  the $\PgU$. The uncertainties associated with the background PDF are in the column labeled {\sc bg}, while the signal PDF, the fitter, tracking efficiency, and effects arising from the efficiency binning are combined in the column labeled {add}. Values in parentheses denote the negative part of the asymmetric uncertainty. The luminosity uncertainty of 11\% is not included in the table.\newline}
  \begin{tabular}{c|c|c|c|c|c|c|c|c|c|c} \hline
$|y|$ &         $A$ & $\eff_{\rm{trig,id}}$ &     $S_{p}$ & $A_{\pt}$ & $A_{\rm vtx}$ & $A_{\rm FSR}$ &        t\&p & $\eff_{\JPsi,\PgU}$ & {\sc bg} &        add.\\ 
\hline \multicolumn{1}{c}{} & \multicolumn{10}{l}{\PgUa \quad  $\pt<30\GeVc$ \quad uncertainties are in percent} \\ \hline
0.0-2.0 &$ 0.5$ &$ 7.5\,( 4.6)$ &$ 0.3\,( 0.3)$ & $ 0.6$ & $ 0.7$ & $ 0.7$ & $ 0.0$ & $ 0.9$ & $ 0.5$ &$ 3.0$ \\ 
0.0-0.4 &$ 0.6$ &$ 6.8\,( 4.9)$ &$ 0.4\,( 0.4)$ & $ 0.7$ & $ 0.3$ & $ 0.7$ & $ 0.6$ & $ 1.5$ & $ 0.1$ &$ 3.0$ \\ 
0.4-0.8 &$ 0.6$ &$ 6.8\,( 4.7)$ &$ 0.4\,( 0.4)$ & $ 0.6$ & $ 0.3$ & $ 0.7$ & $ 0.3$ & $ 1.1$ & $ 5.4$ &$ 3.0$ \\ 
0.8-1.2 &$ 0.5$ &$ 7.5\,( 4.9)$ &$ 0.3\,( 0.3)$ & $ 0.6$ & $ 1.0$ & $ 0.7$ & $ 0.1$ & $ 0.7$ & $ 2.9$ &$ 3.0$ \\ 
1.2-1.6 &$ 0.5$ &$ 7.7\,( 4.0)$ &$ 0.2\,( 0.2)$ & $ 0.6$ & $ 1.2$ & $ 0.6$ & $ 0.2$ & $ 0.5$ & $ 4.0$ &$ 3.0$ \\ 
1.6-2.0 &$ 0.6$ &$ 9.3\,( 4.0)$ &$ 0.0\,( 0.1)$ & $ 0.6$ & $ 0.9$ & $ 0.6$ & $ 0.9$ & $ 0.6$ & $ 5.0$ &$ 3.0$ \\ 
\hline
\end{tabular}

  \label{tab:syst-xsec-rapdiff}
\end{table*}

\section{Results and discussion}
\label{summary}

The analysis of the collision data acquired by the CMS experiment at $\sqrts=7\TeV$,
corresponding to an integrated luminosity of $3.1\pm 0.3\pbinv$,
yields a measurement of the $\PgU(nS)$ integrated production cross sections for the range $|y|<2$:

\ifthenelse{\boolean{cms@external}}{\begin{widetext}}{}

  \begin{align}
    \nonumber
    \sigma(\Pp\Pp \rightarrow \PgUa X ) \cdot {\cal B} (\PgUa \rightarrow \mu^+
    \mu^-) &= 7.37 \pm 0.13 (\rm stat.)^{+0.61}_{-0.42} (\rm syst.)\pm 0.81 (\rm lumi.)\,\text{nb}\,, \\
    \sigma(\Pp\Pp \rightarrow \PgUb X ) \cdot {\cal B} (\PgUb \rightarrow \mu^+
    \mu^-) &= 1.90 \pm 0.09 (\rm stat.)^{+0.20}_{-0.14} (\rm syst.)\pm 0.24 (\rm lumi.)\,\text{nb}\,, \\
    \sigma(\Pp\Pp \rightarrow \PgUc X ) \cdot {\cal B} (\PgUc \rightarrow \mu^+
    \mu^-) &= 1.02 \pm 0.07 (\rm stat.)^{+0.11}_{-0.08} (\rm syst.) \pm 0.11 (\rm lumi.)\,\text{nb}\,.
  \end{align}

\ifthenelse{\boolean{cms@external}}{\end{widetext}}{}

\begin{figure*}[htp]
  \centering
  {\includegraphics[angle=90,width=0.47\textwidth]{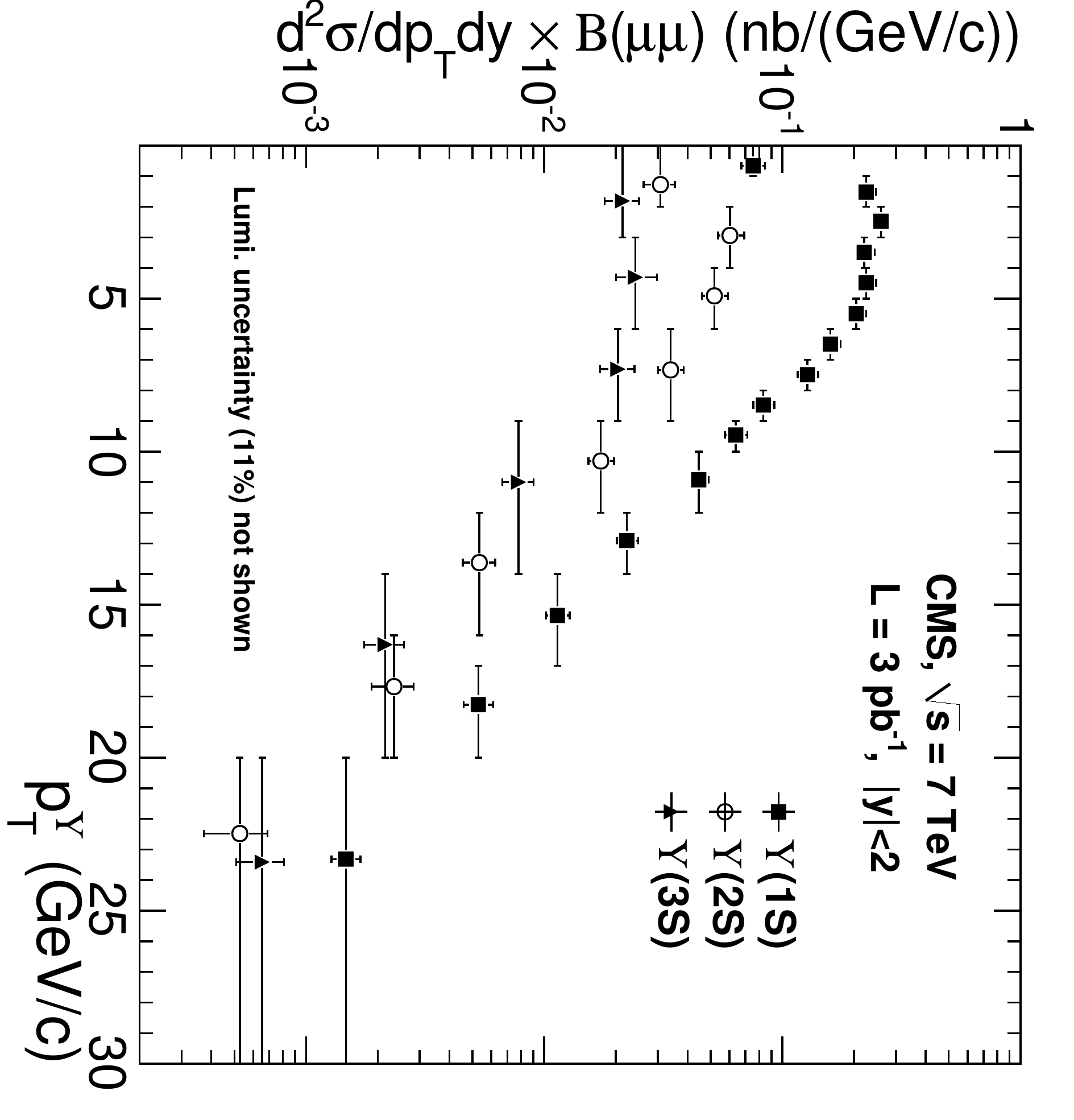}   \label{fig:xsec_overlay}}
  {\includegraphics[angle=90,width=0.47\textwidth]{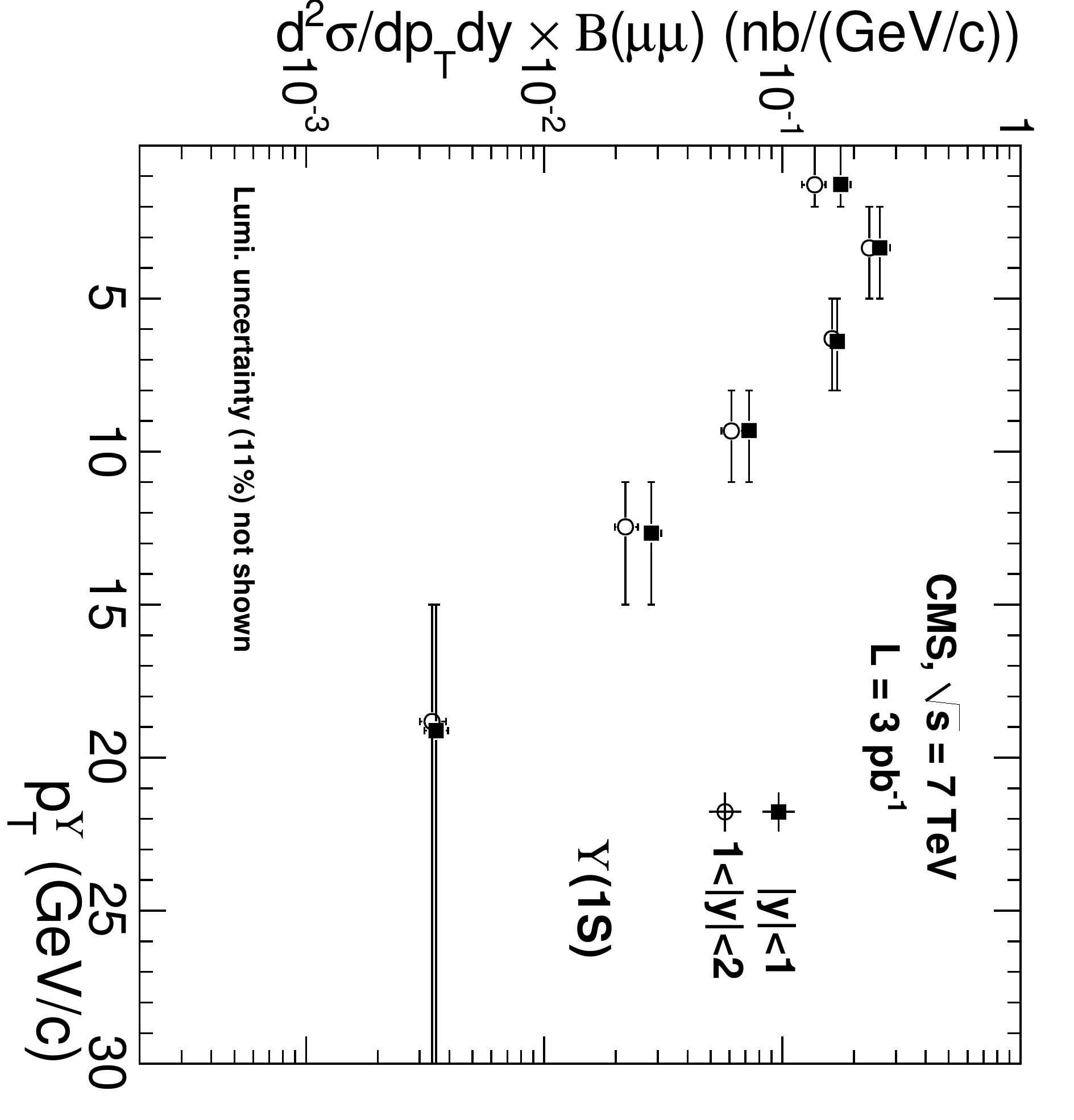}  \label{fig:xsec_1s_2ybin}}
  {\includegraphics[angle=90,width=0.47\textwidth]{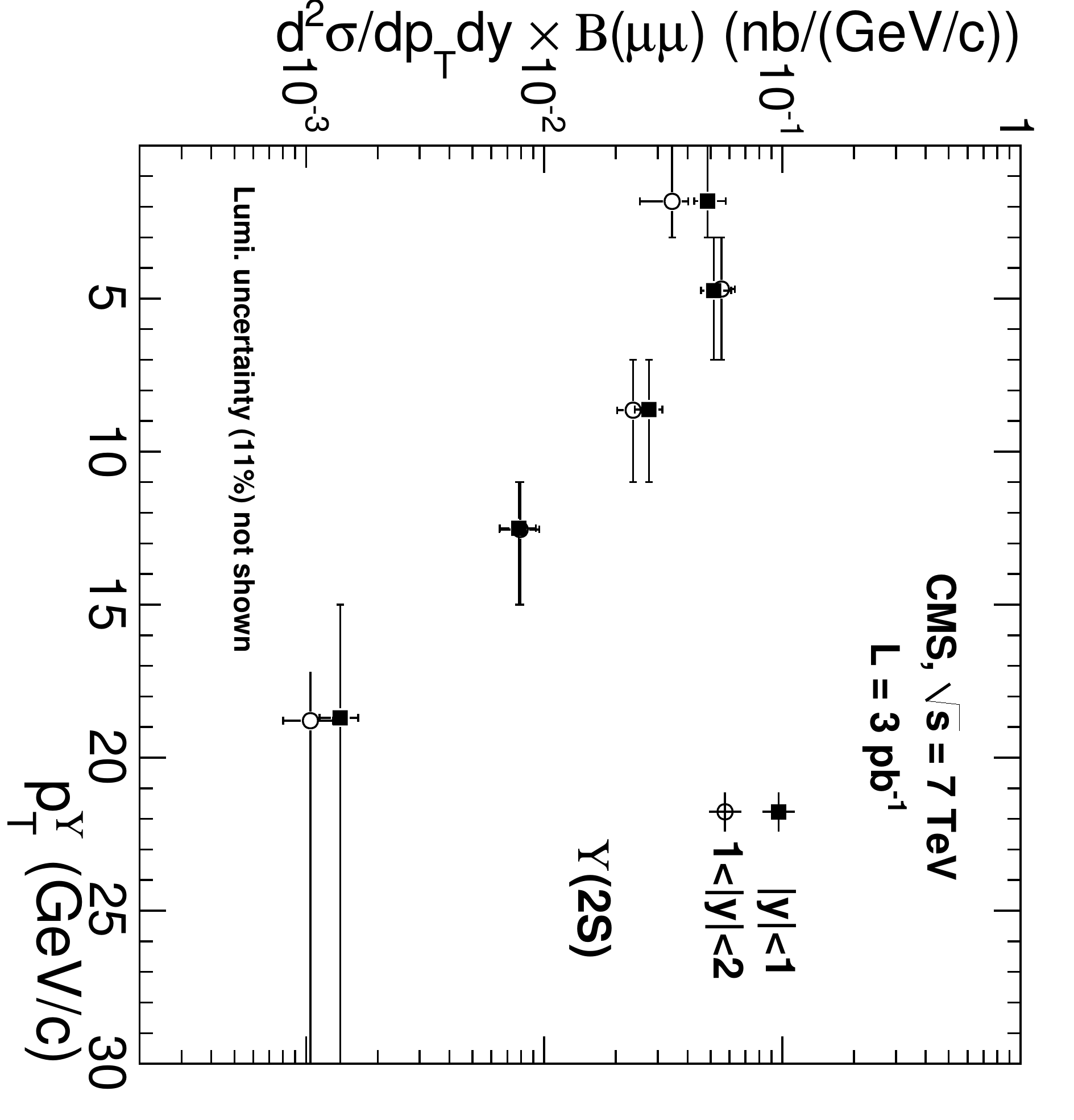}  \label{fig:xsec_2s_2ybin}}
  {\includegraphics[angle=90,width=0.47\textwidth]{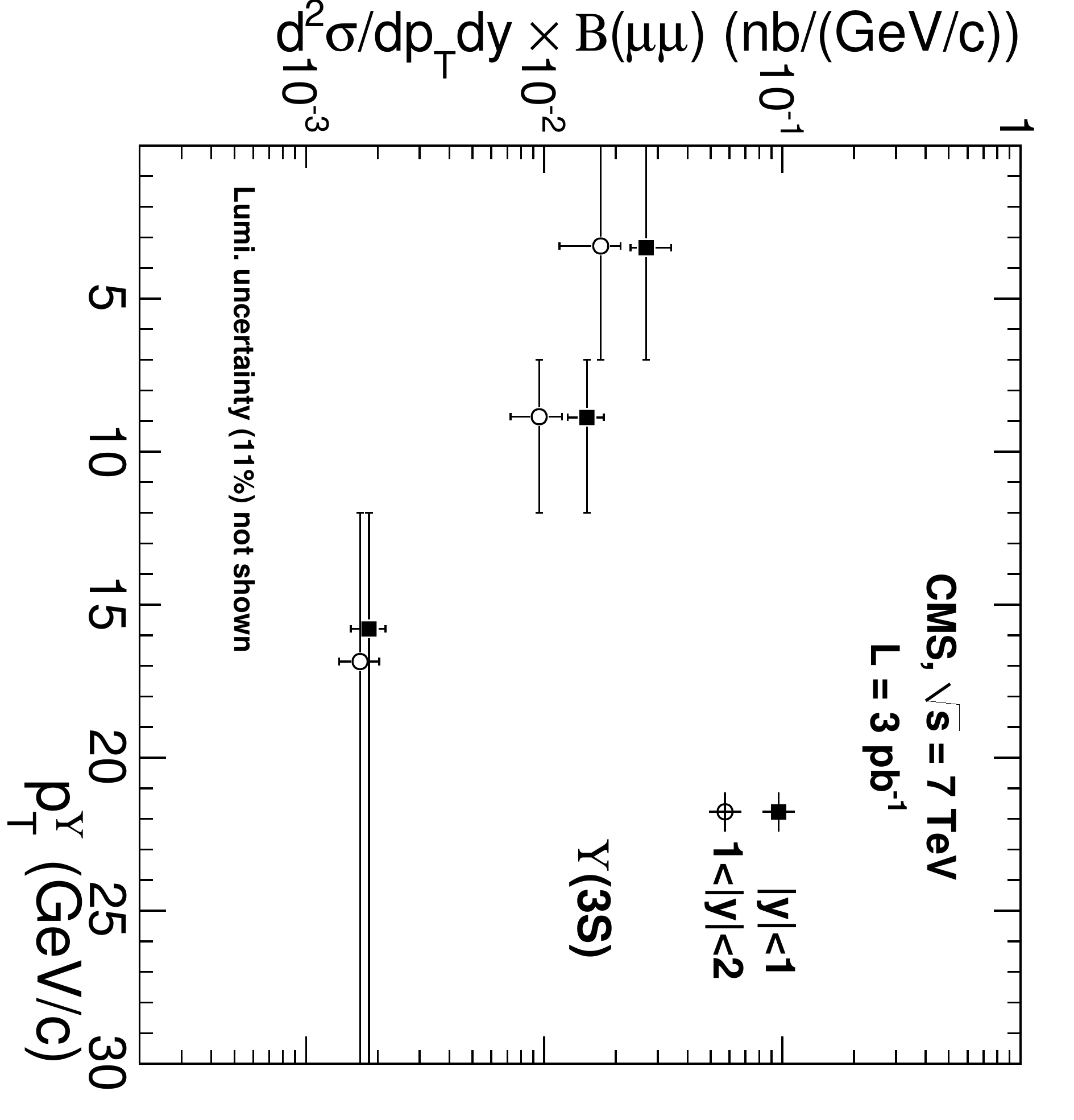}  \label{fig:xsec_3s_2ybin}}
  \caption{$\PgU(nS)$ differential cross sections in the rapidity interval $|y|<2$\, (top left),
    and in the rapidity intervals  $|y|<1$ and $1<|y|<2$ for the $\PgUa$ (top right), $\PgUb$ (botttom left) and $\PgUc$ (bottom right).
    The uncertainties on the points represent the sum of the statistical and systematic uncertainties added in quadrature, excluding the uncertainty on the integrated luminosity (11\%).}
  \label{fig:xsec_ptdiff}
\end{figure*}

\begin{figure*}[htp]
   \centering
  {\includegraphics[angle=90,width=0.47\textwidth]{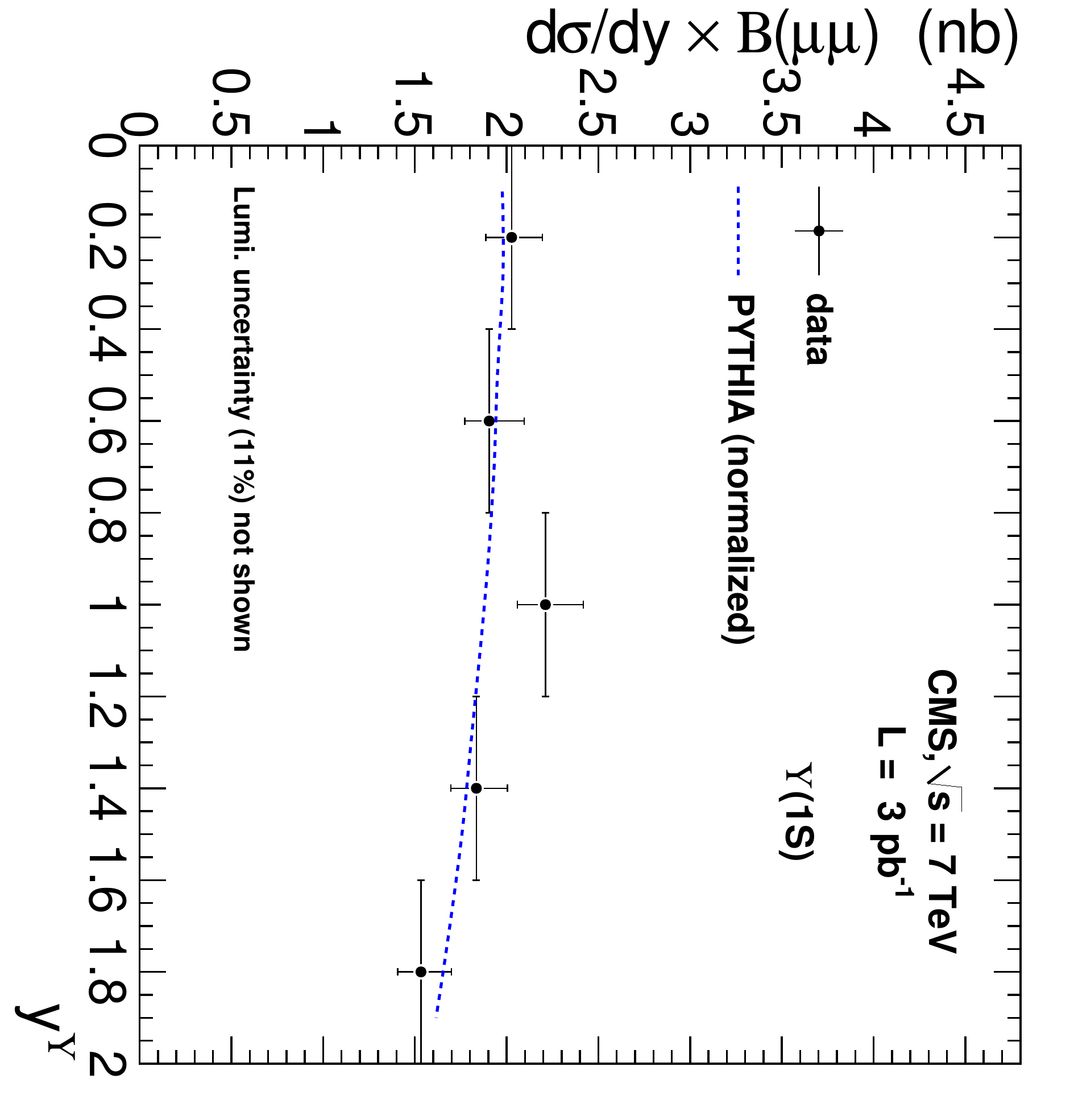}\label{fig:xsec_rapdiff_1s}}
  {\includegraphics[angle=90,width=0.47\textwidth]{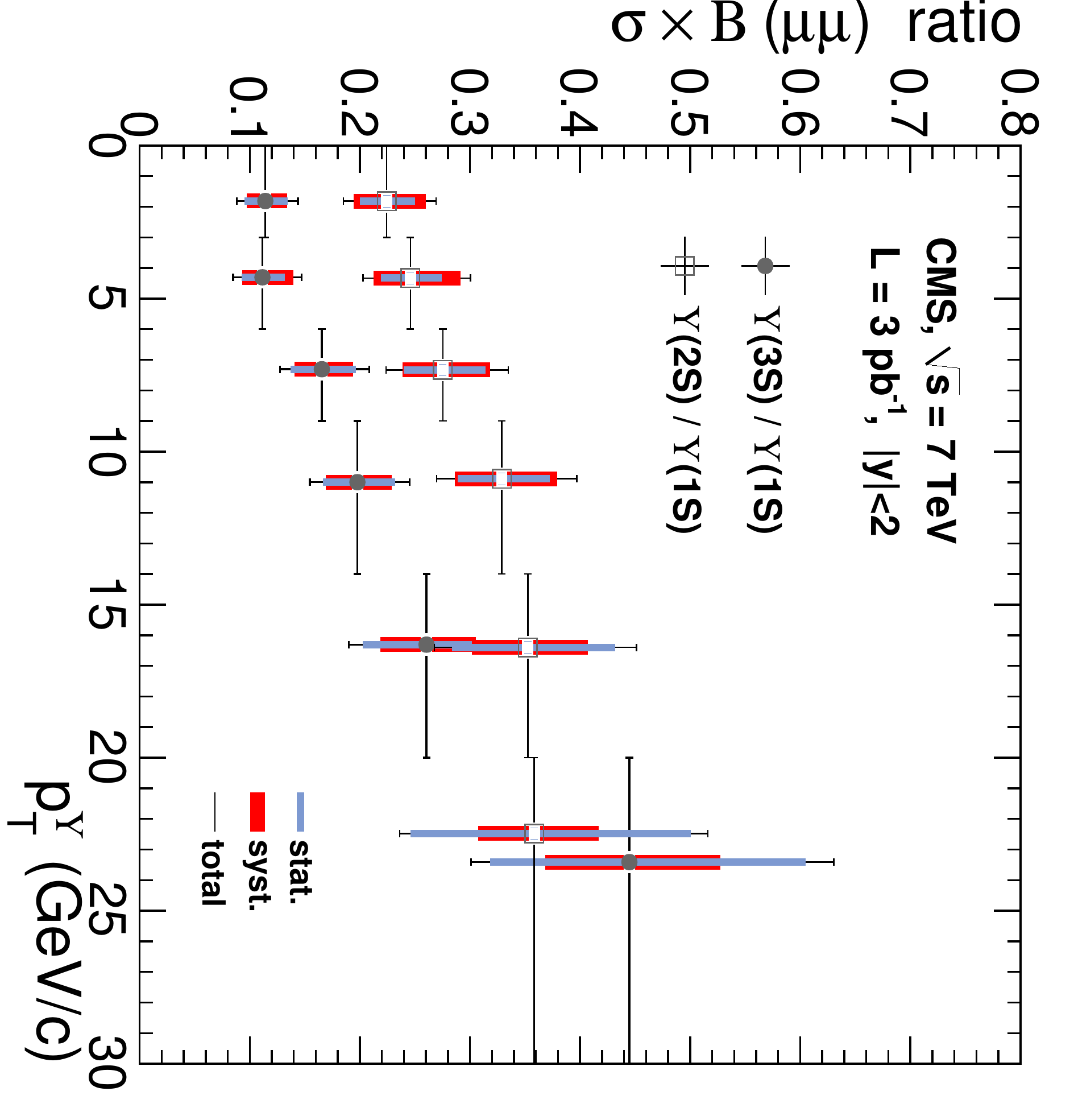}     \label{fig:xsec_ratio}}
  \caption{(Left) $\PgUa$ rapidity-differential cross section in the transverse momentum range $\pt<30$\GeVc (data points) and normalized \PYTHIA prediction (line). The uncertainties on the points represent the sum of the statistical and systematic uncertainties added in quadrature, excluding the uncertainty on the integrated luminosity (11\%).  (Right) $\PgU(nS)$ cross-section ratios as a function of \pt in the rapidity range $|y| < 2$.}
  \label{fig:xsec_rapdiff_ratio}
\end{figure*}

\begin{figure*}[!htp]
  \centering
  {\includegraphics[angle=90,width=0.32\textwidth]{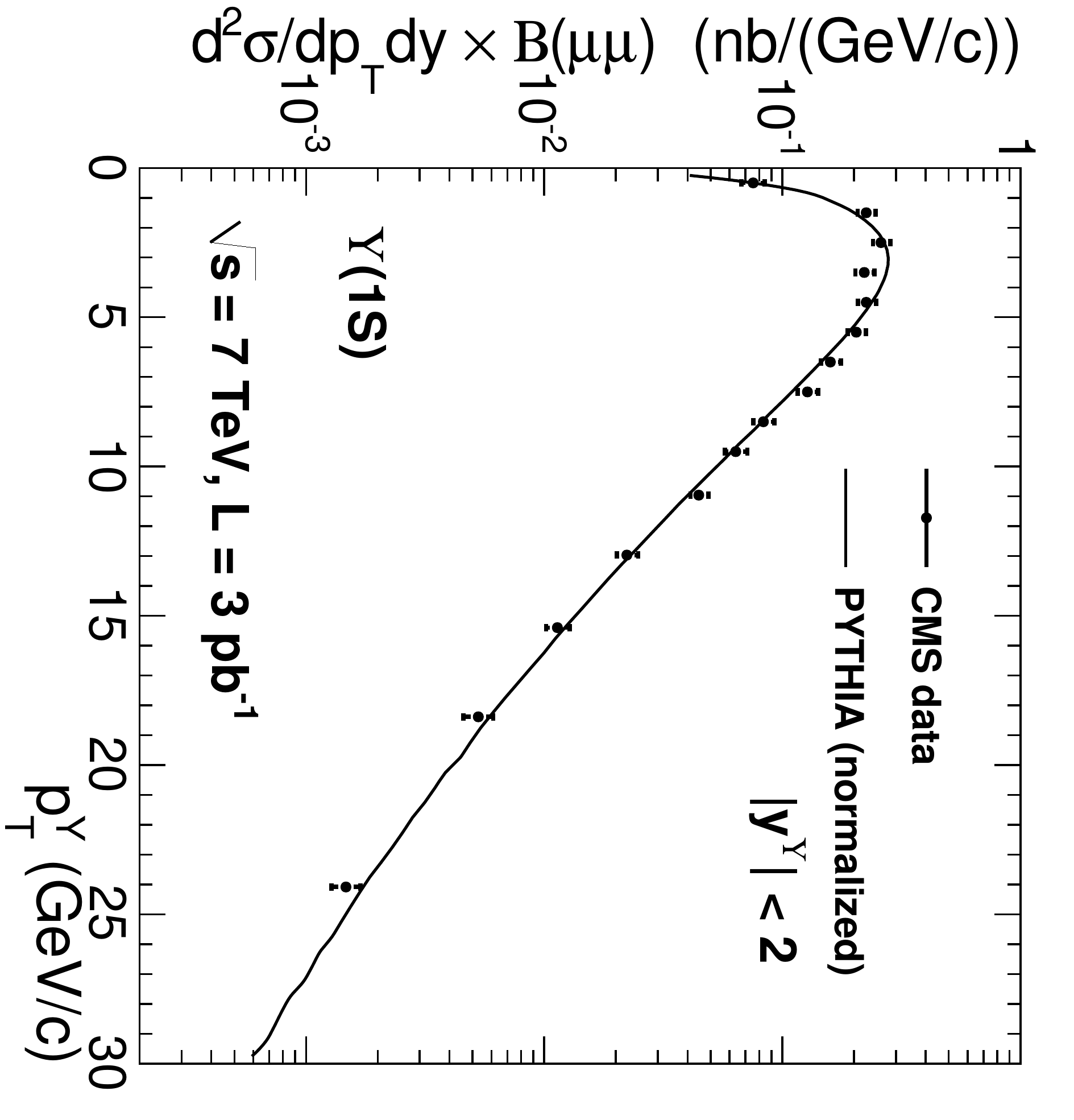} \label{fig:theoryPlot1S}}
  {\includegraphics[angle=90,width=0.32\textwidth]{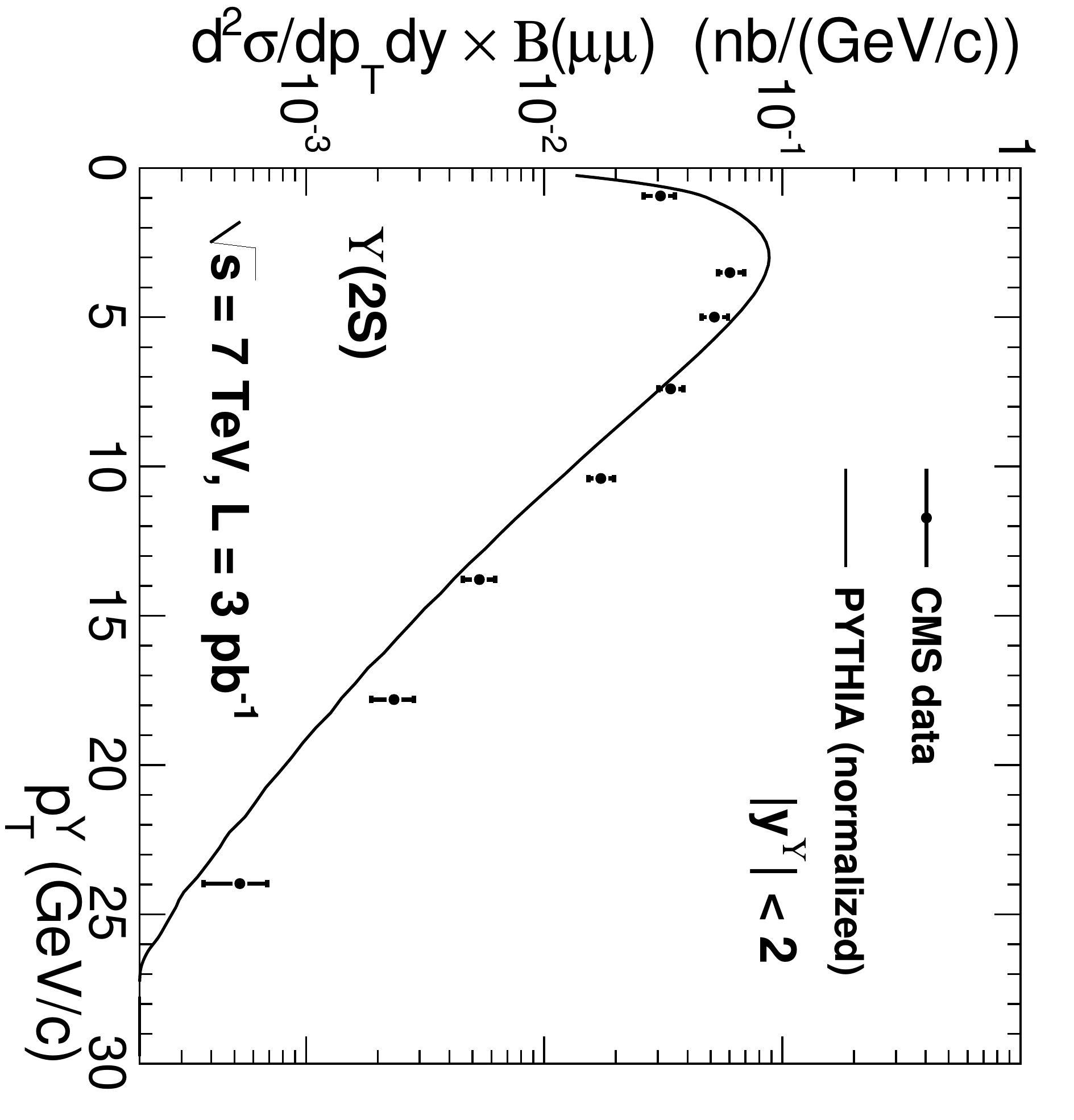} \label{fig:theoryPlot2S}}
  {\includegraphics[angle=90,width=0.32\textwidth]{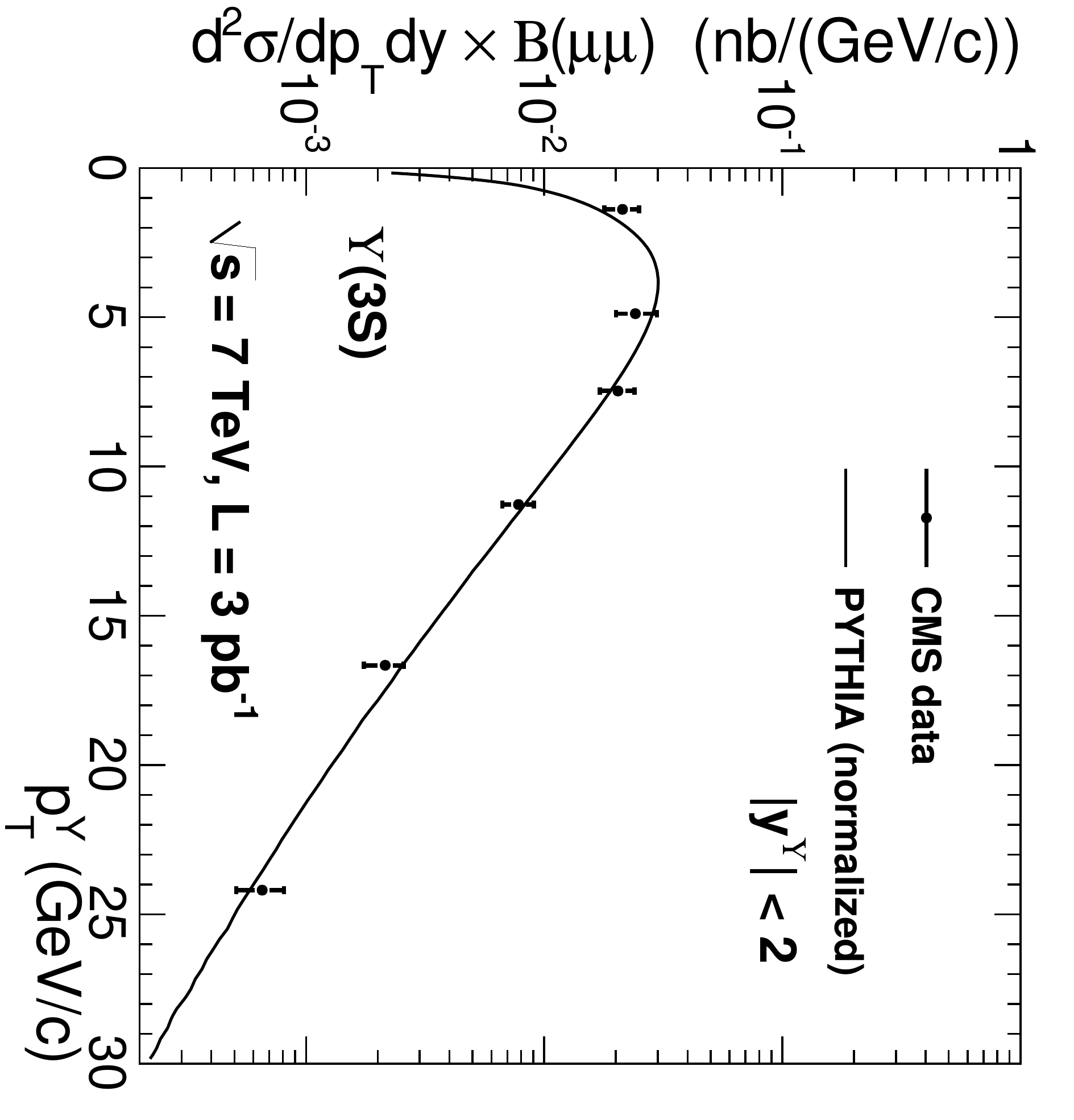} \label{fig:theoryPlot3S}}
  \caption{Differential cross sections of the $\PgU(nS)$ as a function of \pt in the rapidity range $|y|<2$, and
    comparison to the \PYTHIA predictions normalized to the measured \pt-integrated cross sections; $\PgUa$ (left), $\PgUb$ (middle), and $\PgUc$ (right).
    The \PYTHIA curve is used to calculate the abscissa of the data points~\cite{wyatt}.
    The uncertainties on the points represent the sum of the statistical and systematic uncertainties added in quadrature, excluding the uncertainty on the integrated luminosity (11\%).}
  \label{fig:theory}
\end{figure*}

The $\PgUa$ and $\PgUb$ measurements include feed-down from higher-mass states, such as the $\chi_b$ family and the $\PgUc$.
These measurements assume unpolarized $\PgU(nS)$ production.
Assumptions of fully-transverse or fully-longitudinal polarizations change the cross sections by about 20\%.
The \pt-differential $\PgU(nS)$ cross sections for the rapidity intervals $|y|<1$, $1<|y|<2$, and $|y|<2$ are shown in Fig.~\ref{fig:xsec_ptdiff}. 
The \pt dependence of the cross section in the two exclusive rapidity intervals is the same within the uncertainties.
The $\PgUa$ \pt-integrated, rapidity-differential cross sections are shown in the left plot of
Fig.~\ref{fig:xsec_rapdiff_ratio}.
 The cross section shows a slight decline towards $|y|=2$, consistent with the expectation from \PYTHIA.
The ratios of $\PgU(nS)$ cross sections differential in \pt are reported in Table~\ref{tab:xsec_ratio} and shown in the right plot of Fig.~\ref{fig:xsec_rapdiff_ratio}.
The uncertainty associated with the luminosity determination cancels in the computation of the ratios.
Both ratios increase with \pt.
In Fig.~\ref{fig:theory} the differential cross sections for the $\PgUa$, $\PgUb$, and $\PgUc$ are compared to \PYTHIA. 
The normalized \pt-spectrum prediction from \PYTHIA is consistent with the measurements, while the integrated cross section is overestimated by about a factor of two.
We have not included parameter uncertainties in the \PYTHIA calculation.
We do not compare our measurements to other models as no published predictions exist at $\sqrts=7\TeV$ for $\PgU$ production.

\begin{table}[htp]
  \centering
  \caption{The ratios of $\PgU(nS)$ cross sections for different $\PgU$ \pt ranges in the unpolarized scenario. The first uncertainty is statistical and the second is systematic.  The ratios are independent of the luminosity normalization and its uncertainty.\newline}
  \begin{tabular}{ccc} \hline 
\pt (\!\GeVc) & \upsiii / \upsi & \upsii / \upsi \\ \hline
 0--30 & $0.14 \pm 0.01 \pm 0.02$ & $0.26 \pm 0.02 \pm 0.04$ \\ 
 0--3  & $0.11 \pm 0.02 \pm 0.02$ & $0.22 \pm 0.03 \pm 0.04$ \\ 
 3--6  & $0.11 \pm 0.02 \pm 0.03$ & $0.25 \pm 0.03 \pm 0.05$ \\ 
 6--9  & $0.17 \pm 0.03 \pm 0.03$ & $0.28 \pm 0.04 \pm 0.04$ \\ 
 9--14 & $0.20 \pm 0.03 \pm 0.03$ & $0.33 \pm 0.04 \pm 0.05$ \\ 
14--20 & $0.26 \pm 0.07 \pm 0.04$ & $0.35 \pm 0.08 \pm 0.05$ \\ 
20--30 & $0.44 \pm 0.16 \pm 0.08$ & $0.36 \pm 0.14 \pm 0.06$ \\ 
\hline 
 \end{tabular}

  \label{tab:xsec_ratio}
\end{table}

\begin{table}[!h]
  \centering
  \caption{$\PgUa$ cross-section measurements at several center-of-mass collision energies.
    The first uncertainty is statistical, the second is systematic, and the third is associated with the luminosity determination.\newline}
  \begin{tabular}{cccc}
    \hline
    Exp. & $\sqrts$ & $\frac{d\sigma}{dy} (p\overset{\brabar}{p}\to$\PgUa$ X) $ & rapidity \\%
    & (\!\TeV) & $\cdot\;B(\PgU\to\mu\mu)$ & range \\%
    \hline
    CDF & 1.8  & $0.680 \pm 0.015 \pm 0.018 \pm 0.026\,\text{nb}$~\cite{bib-cdfups}  & $|y|<0.4$ \\
    \DZERO  & 1.96 & $0.628 \pm 0.016 \pm 0.065 \pm 0.038~\,\text{nb}$~\cite{bib-d0ups}   & $|y|<0.6$ \\
    CMS & 7.0  & $2.02 \pm 0.03^{+0.16}_{-0.12} \pm 0.22\,\text{nb}$ (this work) & $|y|<1.0$ \\
    \hline
  \end{tabular}
  \label{tab:xsec_vs_sqts}
\end{table}

\begin{figure*}[htp]
  \centering
  \includegraphics[angle=0,width=0.32\textwidth]{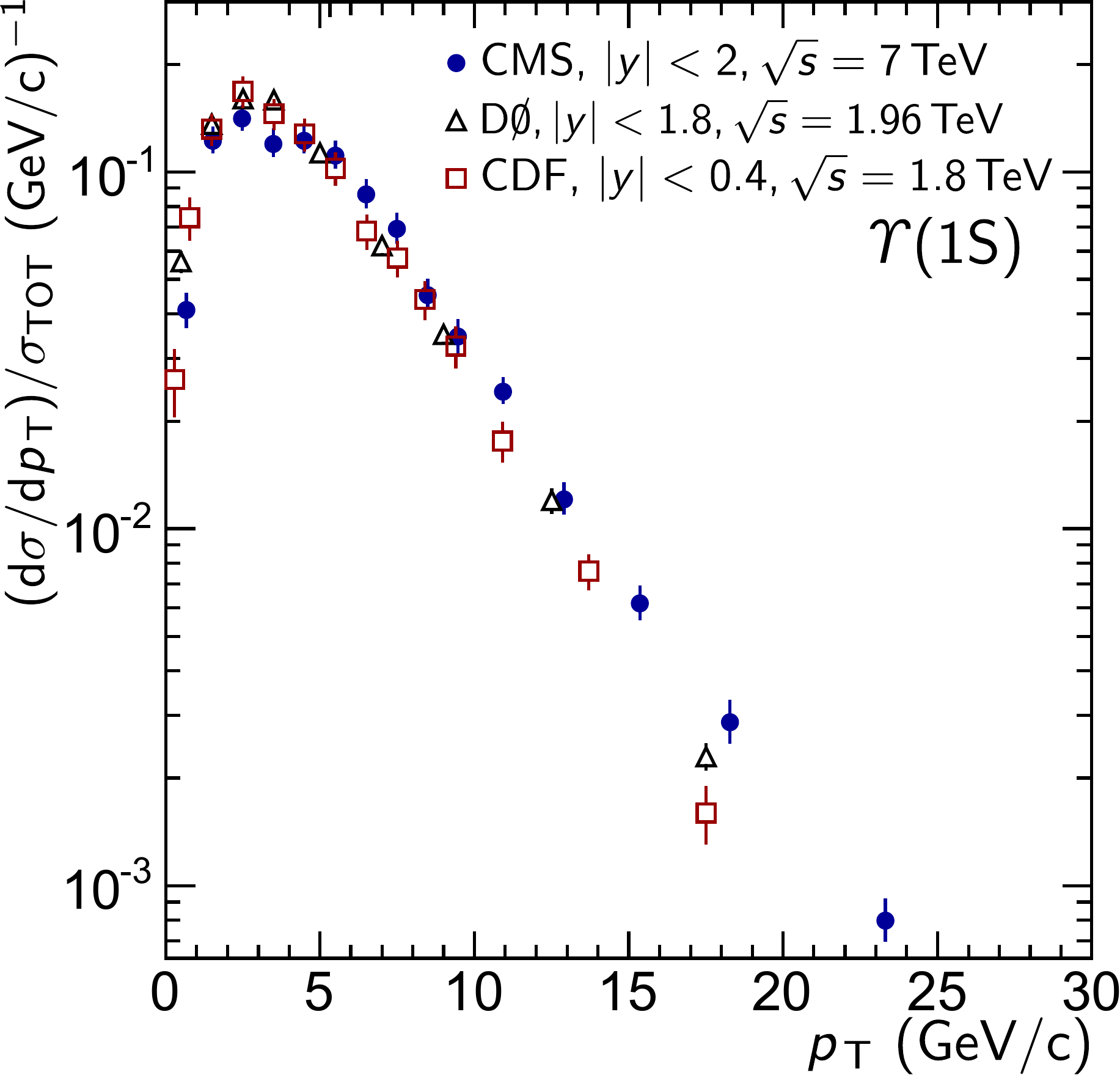} 
  \includegraphics[angle=0,width=0.32\textwidth]{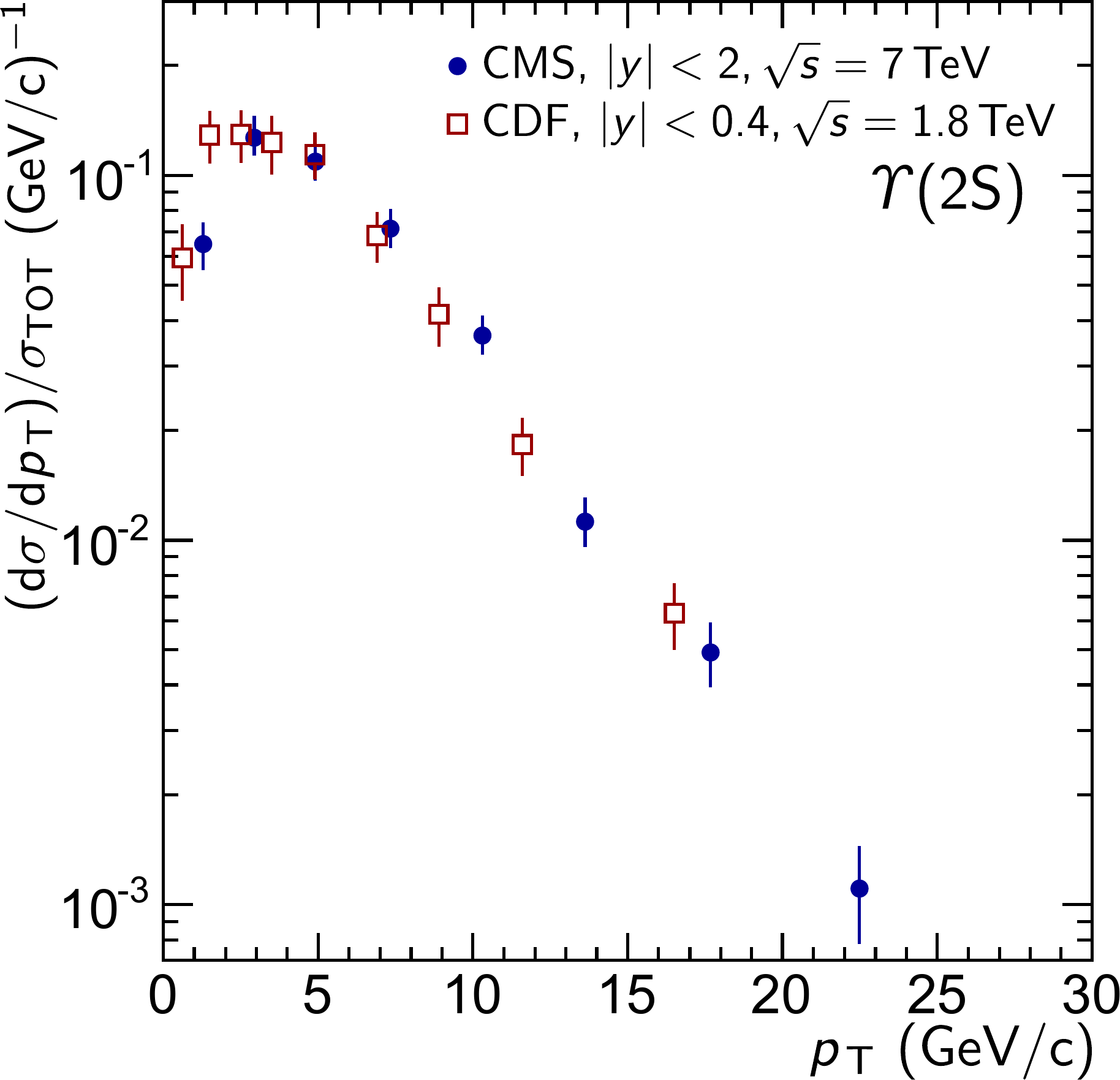} 
  \includegraphics[angle=0,width=0.32\textwidth]{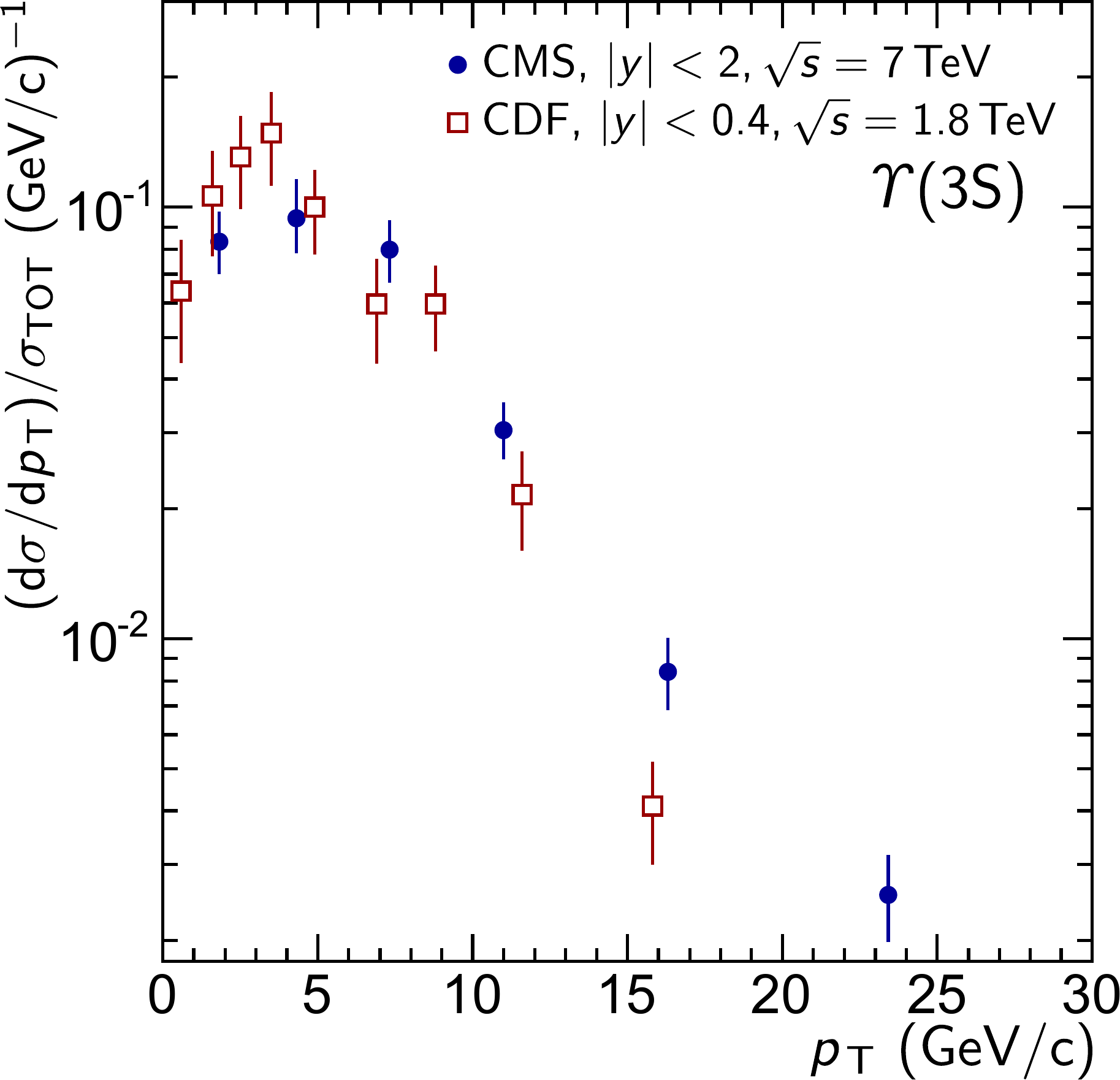} 
  \caption{Comparison of the CMS differential $\PgU(nS)$ cross sections as a function of \pt, normalized by $\sigma_{\rm{TOT}}=\sum (d\sigma/d\pt)\Delta\pt$, to previous measurements; $\PgUa$ (left), $\PgUb$ (middle), and $\PgUc$ (right).}
  \label{fig:xsec_tevatron}
\end{figure*}

The $\PgU(nS)$ integrated cross sections are expected to increase with $\sqrts$.  We compare our measurement of the $\PgUa$ integrated cross section in the central rapidity region $|y|<1$
to previous measurements from the \DZERO and CDF experiments~\cite{bib-d0ups,bib-cdfups} in Table~\ref{tab:xsec_vs_sqts}.
Previous measurements are restricted to the range $\pt < 20$\GeVc and $|y|<0.4$ for CDF and $|y|<1.8$ for \DZERO.
Under the assumption that the cross section is uniform in rapidity for the measurement range of each experiment,
the cross section we measure at $\sqrts=7\TeV$ is about three times larger than the cross section measured at the Tevatron.
Although our measurement extends to higher \pt than the Tevatron measurements, the fraction of the cross section satisfying $\pt > 20$\GeVc is less than 1\% and so can be neglected for this comparison.
We compare the normalized \pt-differential cross sections at the Tevatron to our measurements in Fig.~\ref{fig:xsec_tevatron}.

\section{Summary}
\label{real-summary}

The study of the $\PgU(nS)$ resonances provides important information on the process of hadroproduction of heavy quarks. In this paper we have presented the first measurement of the $\PgU(nS)$ differential production cross section for proton-proton collisions at $\sqrts=7\TeV$.
Integrated over the range $\pt<30\GeVc$ and $|y|<2$, we find
  the product of the $\PgUa$ production cross section and
  dimuon branching fraction to be $\sigma(\Pp\Pp \rightarrow \PgUa X ) \cdot {\cal B} (\PgUa \rightarrow \mu^+\mu^-) =
  7.37 \pm 0.13^{+0.61}_{-0.42}\pm 0.81$\,nb,
where the first uncertainty is statistical, the second is systematic, and the third is associated with the estimation of the integrated luminosity of the data sample.
Under the assumption that the cross section is uniform in rapidity for the measurement range of each experiment,  the cross section we measure at $\sqrts=7\TeV$ is about three times larger than the cross section measured at the Tevatron.
The $\PgUb$ and  $\PgUc$ integrated cross sections and the $\PgUa$, $\PgUb$, and $\PgUc$ differential cross sections in transverse momentum in two regions of rapidity have also been determined.
The differential cross-section measurements have been compared to previous measurements and \PYTHIA.
Finally, the cross section ratios of the three $\PgU(nS)$ have been measured.

The dominant sources of systematic uncertainty on the cross-section measurement arise from the tag-and-probe determination of the efficiencies and from the integrated luminosity normalization.
Both will be reduced with additional data.
The cross sections obtained in this work assume unpolarized $\PgU(nS)$ production.
Assuming fully-transverse or fully-longitudinal polarization changes the cross section by about 20\%.
With a larger accumulated data sample, it will become possible to perform a simultaneous measurement of the polarization and the cross section.
This work provides new experimental results which will serve as input to ongoing theoretical investigations of the correct
description of bottomonium production.

\vspace{5mm}
\noindent
{\bf Acknowledgments}\newline
We wish to congratulate our colleagues in the CERN accelerator departments for the excellent performance of the LHC machine. We thank the technical and administrative staff at CERN and other CMS institutes. This work was supported by the Austrian Federal Ministry of Science and Research; the Belgium Fonds de la Recherche Scientifique, and Fonds voor Wetenschappelijk Onderzoek; the Brazilian Funding Agencies (CNPq, CAPES, FAPERJ, and FAPESP); the Bulgarian Ministry of Education and Science; CERN; the Chinese Academy of Sciences, Ministry of Science and Technology, and National Natural Science Foundation of China; the Colombian Funding Agency (COLCIENCIAS); the Croatian Ministry of Science, Education and Sport; the Research Promotion Foundation, Cyprus; the Estonian Academy of Sciences and NICPB; the Academy of Finland, Finnish Ministry of Education, and Helsinki Institute of Physics; the Institut National de Physique Nucl\'eaire et de Physique des Particules~/~CNRS, and Commissariat \`a l'\'Energie Atomique, France; the Bundesministerium f\"ur Bildung und Forschung, Deutsche Forschungsgemeinschaft, and Helmholtz-Gemeinschaft Deutscher Forschungszentren, Germany; the General Secretariat for Research and Technology, Greece; the National Scientific Research Foundation, and National Office for Research and Technology, Hungary; the Department of Atomic Energy, and Department of Science and Technology, India; the Institute for Studies in Theoretical Physics and Mathematics, Iran; the Science Foundation, Ireland; the Istituto Nazionale di Fisica Nucleare, Italy; the Korean Ministry of Education, Science and Technology and the World Class University program of NRF, Korea; the Lithuanian Academy of Sciences; the Mexican Funding Agencies (CINVESTAV, CONACYT, SEP, and UASLP-FAI); the Pakistan Atomic Energy Commission; the State Commission for Scientific Research, Poland; the Funda\c{c}\~ao para a Ci\^encia e a Tecnologia, Portugal; JINR (Armenia, Belarus, Georgia, Ukraine, Uzbekistan); the Ministry of Science and Technologies of the Russian Federation, and Russian Ministry of Atomic Energy; the Ministry of Science and Technological Development of Serbia; the Ministerio de Ciencia e Innovaci\'on, and Programa Consolider-Ingenio 2010, Spain; the Swiss Funding Agencies (ETH Board, ETH Zurich, PSI, SNF, UniZH, Canton Zurich, and SER); the National Science Council, Taipei; the Scientific and Technical Research Council of Turkey, and Turkish Atomic Energy Authority; the Science and Technology Facilities Council, UK; the US Department of Energy, and the US National Science Foundation.
Individuals have received support from the Marie-Curie IEF program (European Union); the Leventis Foundation; the A. P. Sloan Foundation; the Alexander von Humboldt Foundation; the Associazione per lo Sviluppo Scientifico e Tecnologico del Piemonte (Italy); the Belgian Federal Science Policy Office; the Fonds pour la Formation \`a la Recherche dans l'\'industrie et dans l'\'Agriculture (FRIA-Belgium); and the Agentschap voor Innovatie door Wetenschap en Technologie (IWT-Belgium).

\bibliography{auto_generated}
\cleardoublepage\appendix\section{The CMS Collaboration \label{app:collab}}\begin{sloppypar}\hyphenpenalty=5000\widowpenalty=500\clubpenalty=5000\textbf{Yerevan Physics Institute,  Yerevan,  Armenia}\\*[0pt]
V.~Khachatryan, A.M.~Sirunyan, A.~Tumasyan
\vskip\cmsinstskip
\textbf{Institut f\"{u}r Hochenergiephysik der OeAW,  Wien,  Austria}\\*[0pt]
W.~Adam, T.~Bergauer, M.~Dragicevic, J.~Er\"{o}, C.~Fabjan, M.~Friedl, R.~Fr\"{u}hwirth, V.M.~Ghete, J.~Hammer\cmsAuthorMark{1}, S.~H\"{a}nsel, C.~Hartl, M.~Hoch, N.~H\"{o}rmann, J.~Hrubec, M.~Jeitler, G.~Kasieczka, W.~Kiesenhofer, M.~Krammer, D.~Liko, I.~Mikulec, M.~Pernicka, H.~Rohringer, R.~Sch\"{o}fbeck, J.~Strauss, A.~Taurok, F.~Teischinger, W.~Waltenberger, G.~Walzel, E.~Widl, C.-E.~Wulz
% ...
\vskip\cmsinstskip
% ...
1:~~Also at CERN, European Organization for Nuclear Research, Geneva, Switzerland\\
% ...

\end{sloppypar}
\end{document}